%% file: main.tex
\documentclass[11pt,letterpaper]{article}
\usepackage[T1]{fontenc}
\usepackage{lmodern}
\usepackage[margin=1in]{geometry}
\usepackage{amsmath,amssymb,amsthm}
\usepackage{graphicx,booktabs,tabularx}
\usepackage{algorithm,algpseudocode}
\usepackage[numbers,sort]{natbib}
\usepackage{microtype}
\usepackage[hidelinks]{hyperref}
\hypersetup{pdftitle={Exact Capacitated Vehicle Routing in Base-Two Exponential Time},pdfauthor={Tien Mai}}
\newtheorem{theorem}{Theorem}
\newtheorem{lemma}[theorem]{Lemma}
\newtheorem{proposition}[theorem]{Proposition}
\newtheorem{corollary}[theorem]{Corollary}
\theoremstyle{definition}
\newtheorem{example}[theorem]{Example}
\theoremstyle{remark}
\newtheorem{remark}[theorem]{Remark}
\newcommand{\Os}{O^{*}}
\newcommand{\INF}{+\infty}
\newcommand{\route}{\operatorname{tour}}
\newcommand{\targets}{\mathcal{S}}
\newcommand{\cmax}{c_{\max}}
\newcommand{\Bin}[2]{\binom{#1}{#2}}

\title{Exact Capacitated Vehicle Routing\\in Base-Two Exponential Time}
\author{Tien Mai\\[3pt]\small Singapore Management University}
\date{November 30, 2026}
\begin{document}
\maketitle
\begin{abstract}
We give a deterministic exact algorithm for single-depot capacitated vehicle routing with $n$ customers, interchangeable vehicles, and one capacity constraint. It returns optimal routes or reports infeasibility in $2^n\operatorname{poly}(L_{\mathrm{in}})$ time and space, where $L_{\mathrm{in}}$ is the binary input length. Demands, vehicle capacity $Q$, and directed arc costs may be arbitrary nonnegative integers, and the fleet size may be limited or unlimited. The capacity-indexed dynamic program has a factor linear in $Q$, and bounded-weight subset convolution has a factor linear in the largest arc cost. Our algorithm exploits the fact that covers of a fixed customer set by exactly $b$ routes have the same total unused capacity. Removing a route with at least the average unused capacity allows many customer sets to share one path computation. If $Q>0$ and every demand fits in one vehicle, the bound improves to $O^*(2^{n-(1-\log_3 2)W})$, where $W$ is total demand divided by $Q$. We also give an $O^*(2.70172^n)$ algorithm for adding a feasible route to an arbitrary table of source costs, with extensions to explicitly listed heterogeneous fleets and any fixed number of capacity constraints.
\end{abstract}

\section{Introduction}\label{sec:intro}
The capacitated vehicle routing problem asks for a minimum-cost set of routes that start and end at a depot and together visit every customer exactly once, such that the total demand served on each route does not exceed the vehicle capacity $Q$ \cite{truck,tothvigo}. Every solution combines two decisions: how to partition the customers among the vehicles, and in which order to visit the customers of each part. This paper asks how fast an exact algorithm can make both decisions in the worst case, when the numerical data may be large.

\emph{Two classical bounds.} For a single route, the dynamic program of Bellman \cite{bellman} and Held and Karp \cite{hk} computes the cheapest visiting order of every subset of the $n$ customers in $O(n^22^n)$ time. The cheapest partition then follows by choosing, for every customer set $S$, the customer set $A\subseteq S$ of a last route. Since each customer lies in $A$, in $S\setminus A$, or outside $S$, this takes $\Os(3^n)$ time. Alternatively, routes can be built one customer at a time while the remaining capacity is kept as part of the state \cite{vanhoorn}. This gives base two, but the running time grows linearly with the numerical value of $Q$.

\emph{The question.} When demands, capacity, and costs are encoded in binary, the capacity can be exponential in $n$, and a bound that grows with $Q$ can be far worse than $3^n$. Such a bound also depends on the choice of units: measuring demands a thousand times more finely leaves the feasible solutions unchanged but multiplies the bound by a thousand. Restricting a capacity-indexed table to reachable loads still permits exponentially many load states for a single visited set. If all subset sums of the demands are distinct, as for $q_i=2^{i-1}$, the number of pairs of a set of visited customers and a load of the current route can be of order $3^n$ even when the capacity is binding. With an unlimited fleet and $Q=2^{n-1}$ the count is $2\cdot3^{n-1}+2^{n-1}$, including an empty current route. If the largest customer is unvisited or already on a closed route, each smaller customer may be unvisited, on a closed route, or on the current route. If the largest customer is on the current route, it fills that route, leaving two choices for each smaller customer. Distinct current-route subsets have distinct loads. For subset sum, bin packing, and scheduling, algorithms and lower bounds whose exponential part depends only on the number of items, with all numbers encoded in binary, are an established subject \cite{horowitz,jansen,lente,nederlof}. We ask the corresponding question for vehicle routing:
\begin{quote}
\emph{How much of the gap between $3^n$ and $2^n$ can be closed without any assumption on the magnitudes of the demands, the capacity, and the arc costs?}
\end{quote}
To the best of our knowledge, this is the first deterministic exact algorithm for single-depot CVRP with interchangeable vehicles and one capacity constraint to achieve $2^n\operatorname{poly}(L_{\mathrm{in}})$ time for arbitrary binary-encoded nonnegative integer demands, capacity, and directed arc costs, where $L_{\mathrm{in}}$ is the binary input length. With a single vehicle of sufficient capacity the problem is the asymmetric traveling salesman problem, for which a classical exact algorithm with running time $\Os((2-\varepsilon)^n)$ for a fixed $\varepsilon>0$ remains open \cite{woeginger,miyamoto}. With zero costs and a fleet limit, feasibility alone is bin packing, which can be decided in $\Os(2^n)$ time by inclusion--exclusion \cite{partition}. The weighted versions of that method, however, have running times that grow with the magnitudes of the weights, and the difficulty lies in combining capacities and costs.

\subsection{Main results}\label{sec:general}
Write $\Os$ for a bound up to factors polynomial in the binary input length. In particular, a factor linear in the numerical value of the capacity $Q$ or the largest arc cost is significant.

\begin{theorem}\label{th:base}
Single-depot capacitated vehicle routing with $n$ customers, interchangeable vehicles, one common capacity, and nonnegative binary integer demands and directed arc costs admits a deterministic exact algorithm using $\Os(2^n)$ time and space. For any fleet limit, including an unlimited fleet, it returns optimal routes or reports infeasibility.
\end{theorem}

The algorithm computes the minimum cost for every customer subset and every exact number $b$ of nonempty routes. Once those two quantities are fixed, all feasible covers have the same total unused capacity. For $b\ge1$, some route accounts for at least a $1/b$ share of that total. Removing it leaves at most a fraction $1-1/b$ of the original unused capacity. Several target subsets can therefore use a common threshold in the path computation. Section~\ref{sec:slack} proves that only polynomially many groups are necessary, even when all subset loads are distinct.

Total demand also controls the exponential part. For $Q>0$, an individual demand above $Q$ certifies infeasibility in polynomial time. Otherwise $W=q(V)/Q$ measures total demand in vehicle loads and satisfies $W\le n$. Section~\ref{sec:load} gives a deterministic bound of $\Os(2^{n-(1-\log_3 2)W})$. Thus, when $W$ is at least a fixed positive fraction of $n$, routing admits a fixed base below two for arbitrary directed costs. A separate argument bounds the number of visits to a customer group from the savings obtained by linking two customers directly. Section~\ref{sec:savings} uses this certificate to contract groups. Section~\ref{sec:limits} explains why a method based on three tour fragments can take exponentially more time on the same inputs.

\begin{theorem}\label{th:layercover}
Algorithm~\ref{alg:main}, which builds covers by adding one nonempty route at a time, solves the identical-vehicle problem with or without a fleet limit in $\Os(2.70172^n)$ time and $\Os(2^n)$ space.
\end{theorem}

The layer subroutine of Algorithm~\ref{alg:main} also solves a more general problem. Let $T(A)$ be the minimum cost of a feasible route on $A$, with $T(A)=+\infty$ for an infeasible or empty route. For any source table $f$ with nonnegative integer or infinite entries, running the subroutine over all route sizes computes the minimum of $f(C)+T(S\setminus C)$ over proper subsets $C\subset S$, for every target $S$, within the bounds of Theorem~\ref{th:layercover}. Theorem~\ref{th:conv} states the result by route size as well. Processing one prescribed vehicle at a time therefore permits vehicle-specific costs, capacities and terminals. For the specified batched algorithm the bases $2.70171$ and $2.70172$ bound its worst-case running time from below and above. These bounds concern that algorithm's enumeration of states.

A different choice of the last route gives an intermediate bound for the homogeneous problem. Treating one- and two-route covers separately and removing a smallest route from a cover with at least three routes gives the intermediate base $1+2^{3/4}=2.68179283\ldots$ (Appendix~\ref{app:count}). Removing a route by unused capacity yields Theorem~\ref{th:base}. At the one-vehicle limit, a further uniform improvement to $\Os((2-\varepsilon)^n)$ would apply to weighted asymmetric traveling salesman. Section~\ref{sec:limits} gives the precise computational model in which base two is already tight.

\subsection{Relation to earlier algorithms}\label{sec:related}
Table~\ref{tab:bounds} summarizes the worst-case bounds discussed in this section.

\begin{table}[t]
\centering\small
\caption{Deterministic worst-case running times for capacitated vehicle routing on $n$ customers with identical vehicles.\label{tab:bounds}}
\begin{tabularx}{\linewidth}{@{}Xll@{}}
\toprule
Method & Running time & Numerical dependence\\
\midrule
Enumeration of a last route in every subset & $\Os(3^n)$ & none\\
Capacity-indexed dynamic program \cite{vanhoorn} & $\Os(2^n(\min\{Q,q(V)\}+1))$ & capacity\\
Bounded min-sum subset convolution \cite{convolution} & $\Os(2^n(\cmax+1))$ & arc costs\\
General min-sum subset convolution \cite{stoian} & $\Os(2.9997^n)$ & none\\
Partitions and permutations of the clients \cite{doring} & $2^{O(n\log n)}$ & none\\
Split evaluation alone (Proposition~\ref{pr:splitonly}) & $\Os(2.74946^n)$ & none\\
Arbitrary-source route extension (Theorem~\ref{th:conv}) & $\Os(2.70172^n)$ & none\\
Route-count decomposition (Appendix~\ref{app:count}) & $\Os((1+2^{3/4})^n)$ & none\\
Unused capacity (Theorem~\ref{th:base}) & $\Os(2^n)$ & none\\
Demand-sensitive split (Theorem~\ref{th:load}) & $\Os(2^{n-(1-\log_3 2)W})$ & $W=q(V)/Q$, $Q>0$\\
\bottomrule
\end{tabularx}
\par\smallskip\begin{minipage}{\linewidth}\footnotesize\textit{Notes.} $\Os$ suppresses factors polynomial in the binary input length. In the third column, \emph{none} permits factors polynomial in the bit lengths of the input numbers. The last row assumes $q_i\le Q$ for every customer. Here $q(V)$ is the total demand and $\cmax$ is the largest arc cost. Stoian \cite{stoian} also reports a Las Vegas algorithm with expected running time $\Os(2.9987^n)$. The bound of \cite{doring} is stated for a graph model in which every customer is a client.\end{minipage}
\end{table}

\emph{Exact methods in routing.} The single-depot delivery problem goes back to Dantzig and Ramser \cite{truck}. Laporte, Mercure, and Nobert \cite{laporte} gave an exact branch-and-bound algorithm for the asymmetric capacitated problem. Branch-cut-and-price methods, including the generic solver of Pessoa et al.\ \cite{pessoa}, use route generation, bounds, and instance-dependent search to solve routing problems to optimality. These methods are designed for practical performance. The comparison below concerns worst-case running times as functions of $n$.

\emph{Meet in the middle.} Bidirectional labeling for resource-constrained elementary paths, as in Righini and Salani \cite{righini}, is established in routing. In the exponential-time literature the underlying idea is the meet-in-the-middle scheme of Horowitz and Sahni \cite{horowitz} for knapsack problems, which the Sort \& Search method of Lent\'e et al.\ \cite{lente} extends to scheduling: partial solutions are enumerated in two lists, one list is sorted by a numerical key, and it is searched once for each element of the other. Lemma~\ref{lem:split} applies this scheme to all customer subsets simultaneously, with the load as the key and with the cut chosen to balance the work across all subsets.

\emph{Capacity-indexed dynamic programming.} Van Hoorn \cite{vanhoorn} gives a dynamic program with running time $O((Q+1)(m+1)(n+1)^2 2^n)$ for $m$ vehicles of maximum capacity $Q$. The method accommodates distinct vehicle capacities and start and finish locations. For polynomially bounded $Q$ it yields an $\Os(2^n)$ bound. For interchangeable vehicles one can replace $Q$ by $\min\{Q,q(V)\}$ because a route serves at most the total demand. This capacity factor can still be exponential in $n$.

\emph{Parameterized algorithms.} D\"oring et al.\ \cite{doring} analyze several graph-based vehicle routing problems with respect to parameters including the number of clients and treewidth. For their capacitated variants with a common capacity, they obtain a running time of $k^{O(k)}$ times a polynomial in the graph size, where $k$ is the number of clients \cite[Theorem~4.1]{doring}. When every customer is a client, this is $2^{O(n\log n)}$. Their model permits walks through graph vertices, whereas in our model a route uses the given arcs between successive customers, a distinction that matters for nonmetric costs.

\emph{Subset convolution.} For nonempty $A$, let $T_0(A)$ be the minimum cost of a feasible route serving exactly $A$, with $T_0(A)=\INF$ if the load of $A$ exceeds $Q$, and set $T_0(\varnothing)=0$, so that an available vehicle may remain idle. Adding one vehicle to a table $f$ of cover costs amounts to computing $\min_{C\subseteq S}\{f(C)+T_0(S\setminus C)\}$ for all $S$, which is a min-sum subset convolution. A direct evaluation considers $3^n$ pairs. The partition algorithms of Bj\"orklund, Husfeldt, and Koivisto \cite{partition} distinguish feasibility from optimization with weights. For min-sum subset convolution, Bj\"orklund et al.\ \cite{convolution} give $\Os(2^n\Delta)$ time for integer entries bounded in absolute value by $\Delta$. Let $\cmax=\max_{i,j}c_{ij}$ be the largest arc cost. A collection of routes that serves each customer of a set at most once uses at most $2n$ arcs, so every finite value is at most $2n\cmax$. Infinite values may therefore be replaced by $2n\cmax+1$, with each output truncated at this sentinel. This gives $\Os(2^n(\cmax+1))$ time for vehicle routing, even when $Q$ is large. This sentinel is for routing cover costs; an external source table may have larger finite entries. For binary integer weights of arbitrary magnitude, Stoian \cite{stoian} reports a deterministic algorithm taking $\Os(2.9997^n)$ time and a Las Vegas algorithm taking $\Os(2.9987^n)$ time in expectation. Applying each algorithm to the vehicle layers gives the corresponding bound for routing. Our improvement uses both the path structure of $T_0$ and the additive form of the capacity test.

\emph{Approximation.} If an approximation suffices and every route has positive cost, base two is available. For positive weights, Stoian \cite{stoianapprox} computes a $(1+\varepsilon)$-approximate min-sum subset convolution in time $2^n\log(2M)/\varepsilon$, up to factors polynomial in $n$ and polylogarithmic in $1/\varepsilon$, where $M$ is the largest finite weight. That paper notes that this applies to minimum-cost $k$-coloring, which takes $k-1$ convolutions with accuracy of order $\varepsilon/(k-1)$. The same argument applies to the recurrence $F_b=F_{b-1}\star T$ of Section~\ref{sec:familykernel}. Take the one-route term $T(S)$ exactly and exclude empty inputs to the approximate convolution. Each remaining finite term is positive. For $n\ge1$, missing values can be encoded by $4n\cmax+1$ and outputs at least this large restored to infinity: every feasible exact optimum is at most $2n\cmax$, and its accumulated approximation factor is at most two. Each layer is evaluated to relative accuracy $\varepsilon/(2n)$ for $0<\varepsilon\le1$. Over at most $n$ layers the approximation factor is at most $(1+\varepsilon/(2n))^n\le e^{\varepsilon/2}\le1+\varepsilon$. The present paper concerns exact solutions.

\emph{Comparison.} Van Hoorn's bound is already $\Os(2^n)$ for polynomially bounded capacity. Bounded-integer subset convolution gives the same base when costs have polynomial magnitude. Theorem~\ref{th:base} attains it even when both numerical parameters grow exponentially in $n$. These bounds concern worst-case operations. The verification in Section~\ref{sec:verify} tests optimality and reconstruction and does not measure a speed ratio against branch-cut-and-price.

\emph{Organization.} Section~\ref{sec:model} specifies the input and the notation. Sections~\ref{sec:slack} and~\ref{sec:load} prove the bounds for interchangeable vehicles. Sections~\ref{sec:familykernel}--\ref{sec:combine} prove the arbitrary-source bound, and Section~\ref{sec:fleet} develops its extensions. Sections~\ref{sec:savings} and~\ref{sec:limits} describe a structural improvement and its limitations.

\section{The routing problem}\label{sec:model}
\subsection{Problem statement}
Let $V=\{1,\ldots,n\}$ be the set of customers and let $0$ denote the depot. The input consists of nonnegative integer demands $q_i$ for $i\in V$, a nonnegative integer capacity $Q$, nonnegative integer arc costs $c_{ij}$ for all ordered pairs of distinct $i,j\in V\cup\{0\}$, and a fleet limit $K\in\{0,1,2,\ldots\}\cup\{\infty\}$. For $X\subseteq V$, the \emph{load} of $X$ is $q(X)=\sum_{i\in X}q_i$. Arc costs may be asymmetric and may violate the triangle inequality. Throughout, $\subset$ denotes proper inclusion.

A \emph{route} of size $t\ge1$ is a sequence $(0,v_1,\ldots,v_t,0)$ in which $v_1,\ldots,v_t$ are distinct customers. It \emph{serves} the set $\{v_1,\ldots,v_t\}$, its load is the load of this set, and its cost is $c_{0v_1}+\sum_{k=1}^{t-1}c_{v_kv_{k+1}}+c_{v_t0}$. A route is \emph{feasible} if its load is at most $Q$. A \emph{cover} of a customer set $S$ is a collection of feasible routes whose customer sets partition $S$, and its cost is the sum of the costs of its routes. A \emph{solution} is a cover of $V$ with at most $K$ routes. The task is to find a solution of minimum cost or to report that none exists.

For $n=0$ the empty collection is the optimal solution. If some $q_i>Q$, or if $V\ne\varnothing$ and $K=0$, the instance is infeasible. These boundary cases can be decided directly in polynomial time.

\subsection{Model of computation and notation}
We measure time and space on a random-access machine with logarithmic cost: an operation on integers takes time polynomial in their bit length, and a stored integer occupies space proportional to its bit length. Write $L_{\mathrm{in}}$ for the bit length of the routing instance. For an externally supplied table $f$, let $\ell_f$ be the maximum bit length of any finite entry, or one if all entries are infinite. Customer sets are stored as $n$-bit masks. For $n\ge1$, subset loads and differences $bQ-q(S)$ with $0\le b\le n$ are bounded in absolute value by $n\max_i q_i+nQ$. Counts of subsets and of pairs of subsets are at most $2\cdot4^n$. The cost of a collection of routes and paths that together use at most $2n$ arcs is at most $2n\cmax$, where $\cmax$ is the largest arc cost, and for an arbitrary source table, as defined below, a computed value may also be the sum of one of its entries and such a cost. Infinite values are flagged separately. It therefore suffices to count arithmetic operations, comparisons, and stored values, and we write $\Os(g(n))$ for $g(n)\operatorname{poly}(L_{\mathrm{in}})$. For an externally supplied table the factor is $\operatorname{poly}(L_{\mathrm{in}}+\ell_f)$. Reading and storing the $2^n$ table entries are counted explicitly.

We write $H(p)=-p\log_2p-(1-p)\log_2(1-p)$ for the binary entropy function, with $H(0)=H(1)=0$, and use the standard bounds
\begin{equation}\label{eq:entropy}
 \frac{2^{kH(j/k)}}{k+1}\ \le\ \binom kj\ \le\ 2^{kH(j/k)}\qquad(k\ge1,\ 0\le j\le k).
\end{equation}
Table~\ref{tab:notation} lists the main notation.

\begin{table}[t]
\centering\small
\caption{Main notation.\label{tab:notation}}
\begin{tabularx}{\linewidth}{@{}lX@{}}
\toprule
Symbol & Meaning\\
\midrule
$V$, $n$, $0$ & set of customers, number of customers, depot\\
$q_i$, $q(X)$ & demand of customer $i$, load (total demand) of the customer set $X$\\
$Q$, $K$, $K'$ & capacity, fleet limit, $\min\{K,n\}$\\
$c_{ij}$, $\cmax$ & cost of the arc from $i$ to $j$, largest arc cost\\
$P(A,i)$ & cheapest path from the depot through exactly $A$ that ends at $i$\\
$R(B,j)$ & cheapest path from $j$ through exactly $B$ that ends at the depot\\
$\route(A)$ & cheapest route serving exactly $A$, regardless of its load\\
$T(A)$ & $\route(A)$ if $A\ne\varnothing$ and $q(A)\le Q$, and $\INF$ otherwise\\
$E_b(S)$, $F_b(S)$ & minimum cover costs with exactly $b$ and at most $b$ nonempty routes, respectively\\
$D$, $g_b(D)$ & total unused capacity $bQ-q(S)$ and its rounded threshold\\
$W$, $\gamma$ & total demand $q(V)/Q$ for $Q>0$, and $1-\log_3 2$\\
$f$ & source table, an arbitrary assignment of values to customer sets\\
$s$, $t$ & size of a target $S$, size of the added route; a source $C$ has size $s-t$\\
$G^f_{s,t}(S)$ & layer vector: cheapest extension of a source to the target $S$ by one route of size $t$\\
$f\star T$ & route convolution, the minimum of $G^f_{s,t}$ over all route sizes $t$\\
$x$, $\tau$ & relative route size $t/n$, threshold $269/930$\\
\bottomrule
\end{tabularx}
\end{table}

\section{Choosing the last route by unused capacity}\label{sec:slack}
Use $E_b(S)$ for the least cost of covering $S$ with \emph{exactly} $b$ nonempty routes. The tables $F_b$ in the convolution algorithm allow at most $b$ routes. Set $E_0(\varnothing)=0$ and $E_0(S)=+\infty$ for $S\ne\varnothing$. Skip $E_b(S)$ if $|S|<b$ or $q(S)>bQ$. For an admissible target define the unused capacity $D=bQ-q(S)$. Every feasible cover of $S$ with $b$ routes has this same $D$.

\begin{lemma}\label{lem:removable}
Fix $b\ge1$ and $S$ with $|S|\ge b$ and $D=bQ-q(S)\ge0$. Choose any $g$ satisfying $(1-1/b)D\le g\le D$. In computing $E_b(S)$ it suffices to extend finite sources $E_{b-1}(C)$ with $C\subset S$ and $(b-1)Q-q(C)\le g$. For every such source the added route on $S\setminus C$ meets the capacity constraint.
\end{lemma}
\begin{proof}
Write $d=(b-1)Q-q(C)$. The unused capacity of the added route equals $D-d$, which is nonnegative for every accepted source. Conversely, remove a route with at least $D/b$ unused capacity from an optimal cover. Its remaining $b-1$ routes have total unused capacity at most $(1-1/b)D$. Replacing them by a cheapest exact-count cover of the same customers preserves this quantity. The removed route and the new source form an optimal cover included in the restricted minimum. When $D=0$, every route is full and the test accepts exactly the finite sources with $d=0$.\end{proof}

\emph{Sharing thresholds across targets.}
Set $g_b(0)=0$. For $D>0$, put $k=\lfloor\log_2 D\rfloor$, $p=\lceil\log_2 b\rceil$, $j=\max(0,k-p)$, and let $g_b(D)=2^j\lfloor D/2^j\rfloor$. For $j=0$ this equals $D$. Otherwise, $0\le D-g_b(D)<2^{k-p}\le D/b$, as required by Lemma~\ref{lem:removable}. Each interval $[2^k,2^{k+1})$ has at most $2^p\le2b$ distinct thresholds. The integer $D$ has at most $1+\lceil\log_2(bQ+1)\rceil$ relevant bit positions. Thus there are only $O(b(1+\log(bQ+1)))$ shared computations at route count $b$.

For example, if $Q=50$ and three routes carry $50,49,8$, then $D=43$ and $g_3(D)=40$. Removing the last route leaves source slack 1. Removing either of the others leaves 43 or 42. The test retains a sufficient source without keeping every way to choose a last route.

\begin{proof}[Proof of Theorem~\ref{th:base}]
For each threshold $g$, let $H_g(C)=E_{b-1}(C)$ for finite sources with $(b-1)Q-q(C)\le g$, and $H_g(C)=+\infty$ otherwise. Process masks $X$ in increasing cardinality. For $i\in X$, compute
\begin{equation}\label{eq:slackpath}
 P_g(X,i)=\min\left\{H_g(X\setminus\{i\})+c_{0i},\ 
 \min_{j\in X\setminus\{i\}}\{P_g(X\setminus\{i\},j)+c_{ji}\}\right\}.
\end{equation}
Each value is the least eligible source cost plus the cost of a path from the depot to $i$ visiting exactly the customers of $X$ outside that source. This follows by induction. The first term starts the path with $i$. Removing the last vertex of a longer path gives one of the terms in the second minimum. For each $S$ assigned to $g=g_b(D)$ set
\begin{equation}\label{eq:slackclose}
 E_b(S)=\min_{i\in S}\{P_g(S,i)+c_{i0}\}.
\end{equation}
Lemma~\ref{lem:removable} gives feasibility and preserves an optimal choice. If a target has no feasible cover, a finite accepted source and path would construct one, so its entry stays infinite. Induction on $b$ proves exactness. The answer on nonempty $V$ is the least $E_b(V)$ over $1\le b\le B=\min(K,n)$, with $B=n$ for an unlimited fleet.

Algorithm~\ref{alg:slack} summarizes the computation. One pass uses $O(n^2 2^n)$ arithmetic operations. Sorting masks by their threshold and computing subset loads fit in this bound. For $B\ge1$ set $\ell_Q=1+\lceil\log_2(BQ+1)\rceil$. Summing the number of passes over $b$ gives $O(B^2\ell_Q n^2 2^n)$ operations in total. The costs, loads, and masks have bit length polynomial in the input length. Each pass can retain the source mask that attains each path value. Save the winning last-route set for each finite $(b,S)$, trace these choices backwards, and recover a cheapest directed tour on each resulting set using the ordinary path table. The workspace and all saved witnesses use $O((n+B)2^n)$ entries. Zero capacity gives a single threshold per count. The empty instance and an empty available fleet are decided directly.\end{proof}

\begin{algorithm}[tbp]
\caption{Optimal routes by unused capacity (Theorem~\ref{th:base})}\label{alg:slack}
\begin{algorithmic}[1]
\State Handle the boundary cases in Section~\ref{sec:model}. Set $B=\min\{K,n\}$.
\State Compute subset loads and the Held--Karp tour predecessors. Initialize $E_0(\varnothing)=0$ and $E_0(S)=+\infty$ otherwise.
\For{$b=1,\dots,B$}
  \State Initialize $E_b$ to $+\infty$.
  \State Group targets $S$ with $|S|\ge b$ and $q(S)\le bQ$ by $g=g_b(bQ-q(S))$.
  \For{each threshold $g$ occurring in a group}
    \State Form $H_g$ from $E_{b-1}$ and compute shared paths by \eqref{eq:slackpath}.
    \State Close paths for targets in this group by \eqref{eq:slackclose}, saving a minimizing source.
  \EndFor
\EndFor
\State Minimize $E_b(V)$ over $1\le b\le B$. Report infeasibility if every value is infinite.
\State Recover an optimal cover using the saved sources and the tour predecessors.
\end{algorithmic}
\end{algorithm}

The proof uses the freedom to choose which route to remove. With a prescribed next vehicle the removable route can have the wrong costs or capacity. An arbitrary table $f(C)$ need not describe any collection of routes. For these inputs the available bound is Theorem~\ref{th:conv}.

\section{An exponent sensitive to total demand}\label{sec:load}
\begin{theorem}\label{th:load}
Under the hypotheses of Theorem~\ref{th:base}, suppose $Q>0$ and $q_i\le Q$ for every customer. Put $W=q(V)/Q$ and $\gamma=1-\log_3 2$. A deterministic algorithm returns the optimal value and routes or reports infeasibility in $\Os(2^{n-\gamma W})$ time and space.
\end{theorem}
\begin{proof}
For a positive integer $a\le n/2$, take the $a$ heaviest customers as a set $A$. In an optimal cover at most $a$ routes meet $A$. If their union is $X$, then $A\subseteq X$ and $q(X\setminus A)\le aQ-q(A)$. Use the exact-count dynamic program on the two downward-closed families
\[
 \mathcal L=\{Y\subseteq V:q(Y\setminus A)\le aQ-q(A)\},
 \qquad \mathcal R=2^{V\setminus A}.
\]
For every $X\in\mathcal L$ containing $A$, combine the two values at $X$ and $V\setminus X$ over counts summing to at most $K$, with at most $a$ routes on the left. Every candidate is feasible, and the split of an optimum occurs. Equations~\eqref{eq:slackpath}--\eqref{eq:slackclose} use only submasks, so the cost including reconstruction is $\Os(|\mathcal L|+|\mathcal R|)$. Index only masks in the two families, locating predecessors by binary search in sorted lists. Every subset of a winning mask belongs to its family, so reconstructing its routes, even by recomputing its path table, fits the same bound.

Choose $a=\lfloor\gamma W\rfloor$, using Theorem~\ref{th:base} if $a=0$. For $a\ge1$, we have $a\le n/2$ because $W\le n$ and $\gamma<1/2$. At least $\lceil W\rceil$ customers have positive demand, since each contributes at most one vehicle load. Thus $a<W$ implies that the smallest demand $u$ in $A$ is positive. For a uniform random subset $Y$ of $V\setminus A$ let $Z=q(Y)$ and $D_0=q(V)$. Markov's inequality applied to $3^{-Z/u}$ gives
\[
 \Pr[Z\le aQ-q(A)]\le 3^{(aQ-q(A))/u}
              \prod_{i\notin A}\frac{1+3^{-q_i/u}}{2}.
\]
Convexity of $x\mapsto\log((1+3^{-x})/2)$ bounds it by $x\log(2/3)$ on $[0,1]$. Hence the probability is at most
\[
 [3^{aQ}(2/3)^{D_0}2^{-q(A)}]^{1/u}\le 2^{-a},
\]
using $aQ\le D_0\log_3(3/2)$ and $q(A)\ge au$. There are $2^a$ ways to choose anchor bits, so $|\mathcal L|\le 2^{n-a}=|\mathcal R|$. Rounding the exponent costs at most a factor of two. The random subset serves only to count masks.

For an integer-only implementation, test $a$ downwards from $\lfloor n/2\rfloor$. Generate the nonanchor subsets under the load threshold by appending a bit only while feasible. Stop a failed test when their count exceeds $2^{n-2a}$. The first successful $a$ is at least $\lfloor\gamma W\rfloor$ if this number is positive. All earlier trials have smaller enumeration caps. If every positive $a$ fails, use Theorem~\ref{th:base}. In that case $\gamma W<1$, so the fallback costs at most twice the stated exponential factor.\end{proof}

For arbitrary demands and $Q>0$, first check in polynomial time whether any $q_i>Q$. Such a customer certifies infeasibility. Including this check gives $\Os(1+2^{n-\gamma W})$ time and space on all inputs with $Q>0$.

For example, if half the customers have zero demand and half demand $2Q/5$ with $5\mid Q$, then $W=n/5$. The bound is $\Os(1.90025^n)$ despite the presence of long feasible routes. For a fixed $W$, the saving in the exponent is bounded by a constant.
\section{The route convolution for arbitrary source tables}\label{sec:familykernel}
\subsection{The three procedures for arbitrary source tables}\label{sec:overview}
The route-convolution algorithm computes, for every customer set $S$, the cost of a cheapest cover of $S$ by routes, adding one route at a time. The basic step is an \emph{extension layer}. Fix a route size $t$ and a target size $s$. Given the value $f(C)$ of each customer set $C$ of size $s-t$, called a \emph{source}, the layer computes for every customer set $S$ of size $s$, called a \emph{target}, the smallest sum of $f(C)$ and the cost of a feasible route serving exactly $S\setminus C$, over all $C\subset S$. For a single target this is an ordinary minimum over subsets, and the $\Os(3^n)$ algorithm evaluates it separately for every target. Our three procedures compute all targets of a layer together. Which one is used depends on the relative route size $t/n$, as shown in Figure~\ref{fig:bases} in Section~\ref{sec:combine}.

\emph{Long routes: ordered sources (Section~\ref{sec:batch}).} Write $q(X)$ for the total demand, or \emph{load}, of a customer set $X$. The key observation is that the capacity test of the added route, $q(S)-q(C)\le Q$, can be written as $q(C)\ge q(S)-Q$. This compares one number attached to the source with a threshold that depends only on the target. Hence, when the sources are sorted by nonincreasing load, every target accepts an initial segment of one common list. For a group of targets that accept a common initial segment, a single dynamic program over pairs of a customer set and a path endpoint extends all sources in the segment at once. A state records only a customer set and an endpoint, although paths from different sources reach it with different loads. This is valid because every source in the segment is feasible for every target in the group. The resulting bound improves as routes get longer (Proposition~\ref{pr:long}).

\emph{Short routes: split evaluation (Section~\ref{sec:split}).} For a short route we use its visiting order instead. We cut the route after a prescribed number of customers, enumerate the possible first parts across all targets, and store, for each endpoint of a first part, the cheapest admissible first part for each load threshold. The cut balances upper bounds on two multinomial counts. The product of these upper bounds is independent of the target size (Proposition~\ref{pr:short}). The resulting bound improves as routes get shorter (Proposition~\ref{pr:short}).

\emph{Intermediate routes: covering families (Section~\ref{sec:combine}).} The two bounds meet at route size about $0.294n$. To lower their common value, we construct a deterministic family of $3t$-customer sets such that every possible $t$-customer route lies in one of them. Within a member of the family, we fix the customers of the source that lie outside the member and solve a smaller ordered-source problem (Proposition~\ref{pr:restrict}).

\subsection{Path tables}
For each nonempty $A\subseteq V$ and $i\in A$, let $P(A,i)$ be the minimum cost of a path that starts at the depot, visits exactly the customers in $A$, each once, and ends at $i$. It satisfies the recurrence of Bellman \cite{bellman} and Held and Karp \cite{hk},
\begin{equation}\label{eq:hk}
 P(\{i\},i)=c_{0i},\qquad
 P(A,i)=\min_{j\in A\setminus\{i\}}\{P(A\setminus\{i\},j)+c_{ji}\}
 \quad (|A|\ge2).
\end{equation}
Computing the same table with transposed costs gives $R(B,j)$, the minimum cost, in the original direction, of a path that starts at $j\in B$, visits exactly $B$, and ends at the depot. For nonempty $A$ let
\[
 \route(A)=\min_{i\in A}\{P(A,i)+c_{i0}\}
\]
be the minimum cost of a route serving exactly $A$, regardless of its load, and let $T(A)=\route(A)$ if $q(A)\le Q$ and $T(A)=\INF$ otherwise. We use these tables only on nonempty sets. The tables $P$ and $R$ have $O(n2^n)$ entries and are computed with $O(n^22^n)$ arithmetic operations. All procedures below assume that these tables have been computed.

\subsection{Layers and the layered recurrence}
For $b\ge0$ and $S\subseteq V$, let $F_b(S)$ be the minimum cost of a cover of $S$ with at most $b$ routes, with $F_b(S)=\INF$ if there is none. Thus $F_b(\varnothing)=0$ for all $b$, and $F_0(S)=\INF$ for nonempty $S$. The optimal value of the problem is $F_{K'}(V)$ with $K'=\min\{K,n\}$, since a cover has at most $n$ routes.

The tables $F_b$ are computed from layers. A \emph{source table} $f$ assigns a value in $\mathbb{Z}_{\ge0}\cup\{\INF\}$ to each subset of $V$. For integers $1\le t\le s\le n$, the \emph{layer vector} $G^f_{s,t}$ is indexed by the $s$-element sets $S\subseteq V$, called \emph{targets}, and is defined by
\begin{equation}\label{eq:layer}
 G_{s,t}^{f}(S)=
 \min_{\substack{C\subset S,\ |C|=s-t\\ q(C)\ge q(S)-Q}}
       \{f(C)+\route(S\setminus C)\},\qquad |S|=s ,
\end{equation}
where an empty minimum equals $\INF$. The sets $C$ in \eqref{eq:layer} are the \emph{sources}. A pair $(C,S)$ that satisfies the constraints in \eqref{eq:layer} is a \emph{candidate}, and $f(C)+\route(S\setminus C)$ is its \emph{value}. The constraint $q(C)\ge q(S)-Q$ is the capacity test $q(S\setminus C)\le Q$ of the added route, written as a condition on the source. In words, $G^f_{s,t}(S)$ is the cheapest way to reach the target $S$ from a source of size $s-t$ by adding one feasible route of size $t$.

\begin{lemma}[Layered recurrence]\label{lem:recur}
For every $b\ge1$ and every nonempty $S\subseteq V$ with $s=|S|$,
\begin{equation}\label{eq:outer}
 F_b(S)=\min\left\{T(S),\
             \min_{1\le t\le\lfloor s/2\rfloor} G_{s,t}^{F_{b-1}}(S)\right\}.
\end{equation}
\end{lemma}
\begin{proof}
\emph{The right side is at least $F_b(S)$.} Every finite term on the right side is the cost of a cover of $S$ with at most $b$ routes. Indeed, $T(S)$ is the cost of one feasible route. A finite value $F_{b-1}(C)+\route(S\setminus C)$ in \eqref{eq:layer} is attained by a cover of $C$ with at most $b-1$ routes together with a route serving $S\setminus C$, whose load $q(S)-q(C)$ is at most $Q$.

\emph{The right side is at most $F_b(S)$.} Suppose $F_b(S)$ is finite and consider an optimal cover. If it consists of one route, its cost is at least $T(S)$. Otherwise it has at least two routes, so one of them serves a set $A$ with $t=|A|\le\lfloor s/2\rfloor$. The other routes, at most $b-1$ in number, form a cover of $C=S\setminus A$, so their cost is at least $F_{b-1}(C)$. The route serving $A$ is feasible, so $q(C)\ge q(S)-Q$, and its cost is at least $\route(A)$. Thus $F_b(S)\ge G^{F_{b-1}}_{s,t}(S)$.\end{proof}

Evaluating \eqref{eq:outer} for $b=1,\ldots,K'$ uses $O(n^3)$ layer vectors. With an unlimited fleet, $F_b=F_n$ for all $b\ge n$, and one can save a factor $n$: write $F$ for this table in place of both $F_b$ and $F_{b-1}$ in \eqref{eq:outer}, and evaluate the recurrence once, in increasing order of $s$. This is possible because all sources in \eqref{eq:layer} have size $s-t<s$.

\subsection{The route convolution}
The source table in \eqref{eq:layer} need not come from routes. For an arbitrary source table $f$, define the \emph{route convolution} of $f$ with the feasible route cost $T$ by
\begin{equation}\label{eq:conv}
 (f\star T)(S)=\min_{C\subset S}\{f(C)+T(S\setminus C)\}=\min_{1\le t\le s}G^f_{s,t}(S),\qquad \varnothing\ne S\subseteq V,\ s=|S|.
\end{equation}
The second equality groups the sources by size. Set $(f\star T)(\varnothing)=\INF$. With $T(\varnothing)=\INF$, this is a min-sum subset convolution whose second operand is the capacitated route cost. The convolution with the function $T_0$ of Section~\ref{sec:related}, in which a vehicle may remain idle, is $\min\{f,f\star T\}$. In this notation, $F_b=F_{b-1}\star T$ on nonempty sets for every $b\ge1$, because removing any one route from a cover with at most $b$ routes leaves a cover with at most $b-1$ routes. Lemma~\ref{lem:recur} is the refinement that a route serving at most half of the customers can be removed. The following theorem gives the bound for an arbitrary source table.

\begin{theorem}[Route convolution]\label{th:conv}
For every source table $f$, all layer vectors $G^f_{s,t}$ with $1\le t\le s\le n$, and hence the route convolution $f\star T$, can be computed in $\Os(2.70172^n)$ time and $\Os(2^n)$ space, where $\Os$ also suppresses factors polynomial in the bit length of the finite entries of $f$. For each finite entry the algorithm returns a \emph{witness}, namely the customer set $S\setminus C$ of an added route that attains the minimum.
\end{theorem}

Theorem~\ref{th:layercover} follows from Lemma~\ref{lem:recur} and Theorem~\ref{th:conv}. Theorem~\ref{th:conv} is stronger in two respects that are used in Section~\ref{sec:fleet}: the first operand is arbitrary, and every route size $1\le t\le s$ is covered, not only $t\le s/2$.

\emph{Plan of the proof.} For a single target, \eqref{eq:layer} is an ordinary minimum over subsets. The three procedures below compute the whole layer vector at once, and each is efficient in a different range of the relative route size $x=t/n$. Together they cover every route size.
\begin{itemize}
\item Section~\ref{sec:batch}: the ordered-source procedure (Proposition~\ref{pr:long}), used for $x\ge1/3$.
\item Section~\ref{sec:split}: the split evaluation (Proposition~\ref{pr:short}), used for $x\le\tau$, where $\tau=269/930$.
\item Section~\ref{sec:combine}: the covering-family procedure (Proposition~\ref{pr:restrict}), used for $\tau<x<1/3$.
\end{itemize}
Section~\ref{sec:combine} then combines the three bounds and completes the proofs of both theorems.

\section{Long routes: extending an ordered family of sources}\label{sec:batch}
This section proves the bound for long routes. The argument has two parts, which we state separately. Lemma~\ref{lem:order} is a grouping argument that does not refer to vehicle routing. It says that if every target accepts a prefix of one common list of sources, and if all targets can be evaluated against a common prefix at once, then few such group evaluations suffice. Its proof is a block decomposition of the list, a standard device for threshold queries that are known in advance. Lemma~\ref{lem:multi} supplies the group evaluation for routing: a single dynamic program extends many sources at once.

\begin{lemma}[Ordered sources]\label{lem:order}
Let $C_1,\ldots,C_N$ be a list of sources and let $\targets$ be a family of $M$ targets. Each target $S$ has a cutoff $k(S)\in\{0,\ldots,N\}$, and each pair of a source and a target has a score $w(C_i,S)\in\mathbb{Z}\cup\{\INF\}$. Fix a size parameter, and let ``polynomial'' and $\Os$ refer to it. Assume the following.
\begin{enumerate}
\item[(i)] Each cutoff and each individual score can be computed in polynomial time.
\item[(ii)] For every $j\le N$, the prefix minima $\min_{i\le j}w(C_i,S)$ for all $S\in\targets$ can be computed together in $\Os(\Gamma)$ time and $\Os(\Gamma)$ space, where $\Gamma\ge\max\{N,M\}$.
\end{enumerate}
Then the minima
\[
 \min_{i\le k(S)} w(C_i,S),\qquad S\in\targets,
\]
can be computed in $\Os(\Gamma+\sqrt{\Gamma NM})$ time and $\Os(\Gamma)$ space. If the procedures in (i) and (ii) return minimizing sources, then so does the combined procedure.
\end{lemma}
In our application the size parameter is the input length of the routing instance.
\begin{proof}
Empty families are immediate, and an empty minimum equals $\INF$.

\emph{Procedure.} Choose an integer $1\le g\le N$ and let $h=\lceil N/g\rceil$ be the block width. For each target $S$ write $k(S)=j(S)h+r(S)$ with $0\le r(S)<h$. The indices $i\le k(S)$ split into the prefix $i\le j(S)h$ and the remainder $j(S)h<i\le k(S)$, which has fewer than $h$ elements. Evaluate the scores in the remainder directly, using (i). Targets with the same value $j(S)\ge1$ form a group. For each group, call (ii) once on the common prefix of length $j(S)h$ and use the result for the targets of the group. The required minimum is the smaller of the two parts, including when either part is empty.

\emph{Running time.} Since $j(S)h\le k(S)\le N$, we have $j(S)\le\lfloor N/h\rfloor\le g$, so there are at most $g$ groups. Grouping the targets takes polynomial time per target. The total time is therefore $\Os(g\Gamma+M\lceil N/g\rceil)$. Take $g=\max\{1,\lfloor\sqrt{NM/\Gamma}\rfloor\}$. Since $M\le\Gamma$, we have $g\le\sqrt N\le N$. If $NM<\Gamma$, then $g=1$ and the time is $\Os(\Gamma+NM)=\Os(\Gamma)$. Otherwise $\sqrt{NM/\Gamma}/2\le g\le\sqrt{NM/\Gamma}$, so $g\Gamma\le\sqrt{\Gamma NM}$ and $M\lceil N/g\rceil\le 2\sqrt{\Gamma NM}+M$.

\emph{Space.} Process the groups one at a time and retain, for each target, the best value found so far and a minimizing source. This uses $\Os(\Gamma+M)=\Os(\Gamma)$ space.\end{proof}

\begin{lemma}[Shared path computation]\label{lem:multi}
Let $1\le t\le s\le n$, and let $\mathcal P$ be a family of $(s-t)$-element subsets of $V$ with values $f(C)\in\mathbb{Z}_{\ge0}\cup\{\INF\}$. The values
\[
 \min_{C\in\mathcal P,\ C\subset S}\{f(C)+\route(S\setminus C)\},\qquad |S|=s,
\]
can be computed for all $s$-element sets $S\subseteq V$, together with a minimizing $C$ for each finite value, with $O(n^2 2^n)$ arithmetic operations and comparisons and $O(n2^n)$ stored values.
\end{lemma}
\begin{proof}
\emph{Dynamic program.} For $X\subseteq V$ with $s-t<|X|\le s$ and $i\in X$, define $\Pi(X,i)$ as follows. If $|X|=s-t+1$, let $\Pi(X,i)=f(X\setminus\{i\})+c_{0i}$ when $X\setminus\{i\}\in\mathcal P$, and $\Pi(X,i)=\INF$ otherwise. If $|X|>s-t+1$, let
\[
 \Pi(X,i)=\min_{j\in X\setminus\{i\}}\{\Pi(X\setminus\{i\},j)+c_{ji}\}.
\]

\emph{Invariant.} We show by induction on $|X|$ that
\begin{equation}\label{eq:invariant}
 \Pi(X,i)=\min_{C\in\mathcal P,\ C\subseteq X\setminus\{i\}}\{f(C)+P(X\setminus C,i)\}.
\end{equation}
In words, $\Pi(X,i)$ is the cheapest way to start from some source in $\mathcal P$ and extend it by a path from the depot that visits exactly the remaining customers of $X$ and ends at $i$. If $|X|=s-t+1$, the only $(s-t)$-element subset of $X\setminus\{i\}$ is $X\setminus\{i\}$ itself, and $P(\{i\},i)=c_{0i}$. If $|X|>s-t+1$, then $|X\setminus C|\ge2$ for every $C$ in \eqref{eq:invariant}, and \eqref{eq:hk} gives $P(X\setminus C,i)=\min_{j}\{P(X\setminus(C\cup\{i\}),j)+c_{ji}\}$ over $j\in X\setminus(C\cup\{i\})$. The pairs $(C,j)$ with $C\subseteq X\setminus\{i\}$ and $j\in X\setminus(C\cup\{i\})$ are exactly the pairs with $j\in X\setminus\{i\}$ and $C\subseteq (X\setminus\{i\})\setminus\{j\}$. Exchanging the two minima and applying the induction hypothesis to $X\setminus\{i\}$ proves \eqref{eq:invariant}.

\emph{Output.} For $|S|=s$, the pairs $(i,C)$ with $i\in S$ and $C\subseteq S\setminus\{i\}$ are the pairs with $C\subset S$ and $i\in S\setminus C$. Hence
\[
 \min_{i\in S}\{\Pi(S,i)+c_{i0}\}
 =\min_{C\in\mathcal P,\ C\subset S}\Bigl\{f(C)+\min_{i\in S\setminus C}\{P(S\setminus C,i)+c_{i0}\}\Bigr\},
\]
which is the required value. There are at most $n2^n$ states, each with at most $n$ predecessors. Storing with each state the source from which its minimizing value originates gives a minimizing $C$.\end{proof}

\begin{proposition}[Long-route bound]\label{pr:long}
For every source table $f$ and every $1\le t\le s\le n$, the layer vector \eqref{eq:layer} can be computed, with witnesses, in
\begin{equation}\label{eq:long}
 \Os\!\left(2^n+\sqrt{2^n \binom{n}{s-t}\binom ns}\right)
\end{equation}
time and $\Os(2^n)$ space.
\end{proposition}
\begin{proof}
We apply Lemma~\ref{lem:order} with the following data.
\begin{itemize}
\item \emph{Sources.} List the $(s-t)$-element sets $C$ with $f(C)<\INF$ in order of nonincreasing load $q(C)$, breaking ties arbitrarily. There are $N\le\binom{n}{s-t}$ of them.
\item \emph{Targets.} The targets are the $M=\binom ns$ sets of size $s$.
\item \emph{Cutoffs.} For a target $S$, the sources with $q(C)\ge q(S)-Q$ form a prefix of the list, including every tie at the threshold. Its length $k(S)$ is found by binary search.
\item \emph{Scores.} Set $w(C,S)=f(C)+\route(S\setminus C)$ if $C\subset S$, and $w(C,S)=\INF$ otherwise. A single score is a lookup in the table of $\route$. By \eqref{eq:layer}, $\min_{i\le k(S)}w(C_i,S)=G^f_{s,t}(S)$.
\item \emph{Prefix minima.} For a prefix $C_1,\ldots,C_j$, Lemma~\ref{lem:multi} with $\mathcal P=\{C_1,\ldots,C_j\}$ computes $\min_{i\le j}w(C_i,S)$ for all $s$-element sets $S$ in $\Os(2^n)$ time and space.
\end{itemize}
Lemma~\ref{lem:order} with $\Gamma=2^n\ge\max\{N,M\}$ gives the bound. Sorting the sources and storing the lists fit within these bounds. The minimizing source $C$ gives the witness $S\setminus C$.\end{proof}

The ordering is what makes the merging in Lemma~\ref{lem:multi} valid. That dynamic program keeps, at each state $(X,i)$, only the cheapest path, and it discards paths from other sources that may carry a different load. Every prefix used for a group passes the capacity test for every target in that group, so the load of the source no longer matters. Containment $C\subset S$ is enforced by the state transitions.

\begin{example}\label{ex:order}
Number five customers in order of decreasing demand, and let their demands be $5,4,3,2,1$ and the capacity be $8$. Take $s=4$ and $t=2$, and suppose that the sources with finite value are the six sets
\[
 \{1,2\},\quad \{1,3\},\quad \{1,4\},\quad \{2,4\},\quad \{2,5\},\quad \{4,5\},
\]
with respective loads $9,8,7,6,5,3$. A source with load $9$ is allowed, since it represents customers covered on earlier routes. The target $S=\{1,2,4,5\}$ has load $12$, so it accepts the sources with load at least $4$, and its cutoff is five. With block width two, one shared path computation uses the first four sources, and the fifth source $\{2,5\}$ is evaluated directly. The source $\{1,3\}$ passes the capacity test but cannot reach $S$, because the path transitions only add customers. Any other target with cutoff four or five uses the same shared path computation.
\end{example}

\section{Short routes: splitting a route of prescribed size}\label{sec:split}
For a short route we keep the order in which its customers are visited and cut the route into a first part and a last part. Fix $1\le t\le s\le n$ and a \emph{split size} $d\in\{0,\ldots,t\}$, the number of customers in the last part. For $0<d<t$, consider a candidate $(C,S)$ and a route serving $S\setminus C$. Let $A$ be the set of the first $t-d$ customers of the route and $B$ the set of its last $d$ customers, and put $L=C\cup A$, so that $|L|=s-d$. Once $L$ is fixed, every possible first part $A$ lies in $L$, whereas every possible last part $B$ lies outside $L$. This gives independent lists of first parts and last parts, coupled only by the load inequality $q(A)+q(B)\le Q$ and a single connecting arc.

\begin{lemma}[Split evaluation]\label{lem:split}
For every source table $f$, every $1\le t\le s\le n$, and every $0\le d\le t$, the layer vector \eqref{eq:layer} can be computed, with witnesses, in $\Os(2^n+X_d+Y_d)$ time and $\Os(2^n)$ space, where
\begin{equation}\label{eq:xy}
 X_d=\binom{n}{s-d}\binom{s-d}{t-d},\qquad
 Y_d=\binom{n}{s-d}\binom{n-s+d}{d}.
\end{equation}
\end{lemma}
The term $2^n$ accounts for the path tables. Given these tables, the procedure takes $\Os(X_d+Y_d)$ time.
\begin{proof}
\emph{Procedure for $0<d<t$.} Initialize every entry of the layer vector to $\INF$, and process each set $L\subseteq V$ with $|L|=s-d$ as follows.
\begin{enumerate}
\item[(1)] For each endpoint $i\in L$, build a list $\mathcal L_i$ that contains, for each $A\subseteq L$ with $|A|=t-d$ and $i\in A$, the triple
\[
       \bigl(q(A),\ f(L\setminus A)+P(A,i),\ A\bigr).
\]
Sort each list by the first coordinate, and store the running minimum of the second coordinate together with a minimizing set.
\item[(2)] For each $d$-element set $B\subseteq V\setminus L$ and each $i\in L$, let $\varphi_i(B)$ be the minimum of the second coordinate over the entries of $\mathcal L_i$ with $q(A)\le Q-q(B)$, found by binary search, with $\varphi_i(B)=\INF$ if there is none. Let
\[
 \psi_i(B)=\min_{j\in B}\{c_{ij}+R(B,j)\},
\]
and replace the entry for the target $S=L\cup B$ by the minimum of its current value and $\varphi_i(B)+\psi_i(B)$.
\end{enumerate}

\emph{Every computed value is the value of a route.} Each finite value $\varphi_i(B)+\psi_i(B)$ is obtained from some $A\subseteq L$ with $|A|=t-d$, $i\in A$, and $q(A)+q(B)\le Q$, and some $j\in B$. Put $C=L\setminus A$. The value equals $f(C)$ plus the cost of a route that follows a cheapest path from the depot through $A$ to $i$, then the arc from $i$ to $j$, and then a cheapest path from $j$ through $B$ to the depot. This route serves exactly $A\cup B=S\setminus C$, so its cost is at least $\route(S\setminus C)$. Moreover $C\subset S$, $|C|=s-t$, and $q(S)-q(C)\le Q$, so $(C,S)$ is a candidate and the computed value is at least $G^f_{s,t}(S)$.

\emph{Every optimal candidate is found.} Conversely, let $(C,S)$ be a candidate that attains the minimum in \eqref{eq:layer}, and let $(0,v_1,\ldots,v_t,0)$ be a cheapest route serving $S\setminus C$. Put $A=\{v_1,\ldots,v_{t-d}\}$, $B=\{v_{t-d+1},\ldots,v_t\}$, $i=v_{t-d}$, $j=v_{t-d+1}$, and $L=C\cup A$. The cost of the route is at least $P(A,i)+c_{ij}+R(B,j)$. When the procedure processes $L$, $B$, and $i$, the entry for $A$ lies in $\mathcal L_i$ and satisfies $q(A)\le Q-q(B)$, so $\varphi_i(B)\le f(C)+P(A,i)$ and $\psi_i(B)\le c_{ij}+R(B,j)$. Hence the computed entry for $S$ equals $G^f_{s,t}(S)$, and the recorded sets $A$ and $B$ give the witness $A\cup B$.

\emph{Running time and space.} Across all $L$, the number of first parts processed is $X_d$ and the number of last parts is $Y_d$, each with at most $n$ endpoints. Enumerating sets of a given size takes polynomial time per generated set, and sorting and binary searches add polynomial factors. Processing one $L$ at a time keeps the storage within $\Os(2^n)$.

\emph{The cases $d=0$ and $d=t$.} Here the procedure instead enumerates, for every target $S$, all $t$-element subsets $A\subseteq S$, and for those with $q(A)\le Q$ it evaluates $f(S\setminus A)+\route(A)$ using the table of $\route$. The number of pairs $(S,A)$ is $\binom ns\binom st=X_0=Y_t$.\end{proof}

The next proposition chooses the split size. Recall that $H$ is the binary entropy function.

\begin{proposition}[Short-route bound]\label{pr:short}
Fix constants $0<\alpha<1/2$ and $\kappa\ge0$. For every source table $f$ and all $1\le t\le s\le n$ with $t\le\alpha n+\kappa$, choosing $d\in\{0,\ldots,t\}$ to minimize $X_d+Y_d$ in Lemma~\ref{lem:split} computes the layer vector \eqref{eq:layer} in time
\begin{equation}\label{eq:short}
                  \Os\!\left(2^{n(1+H(\alpha)/2)}\right)
\end{equation}
and space $\Os(2^n)$.
\end{proposition}
\begin{proof}
\emph{Two upper bounds.} The two factors in each product of \eqref{eq:xy} form a multinomial coefficient. For $X_d$ the cell counts are $(t-d,s-t,n-s+d)$, and for $Y_d$ they are $(d,n-s,s-d)$. Under the vector of cell probabilities $(\alpha/2,(1-\alpha)/2,1/2)$, the probability of either count vector is at most one. Consequently
\begin{align*}
 X_d &\le E_1(d)=2^n\alpha^{-(t-d)}(1-\alpha)^{-(s-t)},\\
 Y_d &\le E_2(d)=2^n\alpha^{-d}(1-\alpha)^{-(n-s)}.
\end{align*}

\emph{Balancing.} As functions of a real variable $d$, $E_1$ is decreasing and $E_2$ is increasing. Their product is $4^n\alpha^{-t}(1-\alpha)^{-(n-t)}$, which does not depend on $d$ or $s$. Let $\Lambda$ be its square root, and note that $\Lambda\ge2^n$. At the real point $d^*$ where $E_1(d^*)=E_2(d^*)$, both bounds equal $\Lambda$. There are three cases.
\begin{itemize}
\item If $d^*\in[0,t]$, then $d=\lfloor d^*\rfloor$ satisfies $Y_d\le E_2(d)\le\Lambda$ and $X_d\le E_1(d)\le\Lambda/\alpha$.
\item If $d^*<0$, then $X_0\le E_1(0)\le\Lambda$ and $Y_0=\binom ns\le 2^n\le\Lambda$.
\item If $d^*>t$, then $Y_t\le E_2(t)\le\Lambda$ and $X_t=\binom{n}{s-t}\le 2^n\le\Lambda$.
\end{itemize}
In every case $\min_d(X_d+Y_d)\le2\Lambda/\alpha$. The algorithm finds the minimizing integer $d$ by computing the binomial coefficients in \eqref{eq:xy} exactly.

\emph{Dependence on $t$.} As a function of $t$, $\Lambda=2^n(1-\alpha)^{-n/2}((1-\alpha)/\alpha)^{t/2}$ increases, because $\alpha<1/2$. At $t=\alpha n$ it equals $2^n(\alpha^{-\alpha}(1-\alpha)^{-(1-\alpha)})^{n/2}=2^{n(1+H(\alpha)/2)}$. Hence $t\le\alpha n+\kappa$ implies $\Lambda\le((1-\alpha)/\alpha)^{\kappa/2}\,2^{n(1+H(\alpha)/2)}$. Since $\Lambda\ge2^n$, Lemma~\ref{lem:split} yields \eqref{eq:short}.\end{proof}

\subsection{What the split evaluation gives by itself}
Rahwan and Jennings \cite{rahwan} restrict the splits evaluated by the dynamic program for complete set partitioning in coalition structure generation; in particular, coalitions with more than about $2n/3$ members, other than the grand coalition, are not split. A related two-thirds restriction on the sizes of the sets combines here with the split evaluation and gives a magnitude-independent benchmark for Theorem~\ref{th:layercover}.

\begin{proposition}[Split-only bound]\label{pr:splitonly}
Using Lemma~\ref{lem:split} as the only layer procedure, capacitated vehicle routing with identical vehicles, with or without a fleet limit, can be solved in $\Os(432^{n/6})$ time and $\Os(2^n)$ space. Here $432^{1/6}\approx2.74946$.
\end{proposition}
\begin{proof}
Let $\eta=\lceil 2n/3\rceil$ and $K'=\min\{K,n\}$, and assume $n\ge1$ and $K\ge1$, since otherwise the answer is immediate.

\emph{Tables on sets of size at most $\eta$.} Compute $F_b(S)$ for all $|S|\le\eta$ and all $b\le K'$ by Lemma~\ref{lem:recur}. This uses only sources of size less than $|S|$. Every layer in \eqref{eq:outer} then has $t\le\lfloor s/2\rfloor\le\lfloor\eta/2\rfloor\le n/3+1/2$. By Proposition~\ref{pr:short} with $\alpha=1/3$ and $\kappa=1/2$, each layer takes $\Os(2^{n(1+H(1/3)/2)})$ time. Since $H(1/3)=\log_23-2/3$, we have $2^{6(1+H(1/3)/2)}=2^6\cdot 3^3\cdot 2^{-2}=432$.

\emph{Combination.} The algorithm returns the minimum of the following two families of values. The first consists of $T(A)+F_{K'-1}(V\setminus A)$ over nonempty $A$ with $|V\setminus A|\le\eta$. The second consists of $F_b(Z)+F_{K'-b}(V\setminus Z)$ over $1\le b<K'$ and $Z$ with $|Z|\le\eta$ and $|V\setminus Z|\le\eta$. Every finite value is the cost of a solution with at most $K'$ routes.

\emph{An optimal solution is among these values.} Consider an optimal solution with routes serving $A_1,\ldots,A_k$, where $k\le K'$. If some $|A_p|\ge n/3$, then $|V\setminus A_p|\le\eta$, and the cost is at least $T(A_p)+F_{K'-1}(V\setminus A_p)$. Otherwise every route serves fewer than $n/3$ customers. Let $Z$ be the union of $A_1,\ldots,A_j$, where $j$ is the smallest index with $|Z|\ge n/3$. Then $|Z|<2n/3$ and $|V\setminus Z|\le 2n/3$, and the cost is at least $F_j(Z)+F_{K'-j}(V\setminus Z)$.

\emph{Routes, time, and space.} With an unlimited fleet the budgets can be dropped. There are $O(n2^n)$ values to compare, and the tables use $\Os(2^n)$ space. Storing a witness with every entry $F_b(S)$, and the sets that attain the final minimum, recovers the routes as in the proof of Theorem~\ref{th:layercover} in Section~\ref{sec:combine}.\end{proof}

\section{Intermediate routes and the proofs of the theorems}\label{sec:combine}
\subsection{The two bounds have opposite dependence on the route size}
\begin{figure}[t]
\centering
\includegraphics[width=0.74\textwidth]{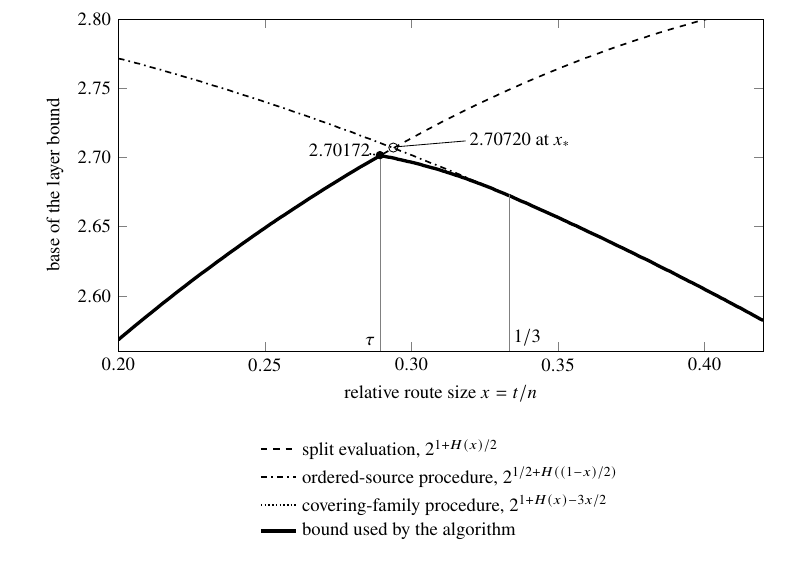}
\caption{Bases of the three layer bounds as functions of the relative route size $x=t/n$.\label{fig:bases} The split evaluation is used for $x\le\tau$, the covering-family procedure for $\tau<x<1/3$, and the ordered-source procedure for $x\ge1/3$. Without the covering families, the two other bounds meet at $x_*\approx0.2938$ with base $2.70720$ (Section~\ref{sec:crossing}).}
\end{figure}

The exponent $1+H(\alpha)/2$ in Proposition~\ref{pr:short} increases with $\alpha$ on $(0,1/2)$. The bound of Proposition~\ref{pr:long} decreases with the route size. To see this, put $x=t/n$ and $z=s/n$. By \eqref{eq:entropy}, the symmetry $H(p)=H(1-p)$, and the concavity of $H$,
\begin{equation}\label{eq:jensen}
 \frac1n\log_2\sqrt{2^n \binom{n}{s-t}\binom ns}
 \ \le\ \tfrac12+\tfrac12\bigl(H(z-x)+H(1-z)\bigr)
 \ \le\ \tfrac12+H\!\left(\frac{1-x}{2}\right).
\end{equation}
The two entropy arguments in the middle expression sum to $1-x$. For fixed $x$, the right side is attained by the middle expression when $z=(1+x)/2$. As $x$ increases, the right side decreases. Figure~\ref{fig:bases} shows both bounds. They meet at $x\approx0.294$, and the covering families of the next subsection lower the maximum.

\subsection{Restricting the possible customers on the last route}
Fix integers $1\le t\le r\le n$. A family $\mathcal U$ of $r$-element subsets of $V$ is a \emph{$(t,r)$-covering family} if every $t$-element subset of $V$ is contained in some member of $\mathcal U$. If the added route has size $t$, its customer set lies in some member $U$ of such a family. The layer can then be computed inside $U$, on $r$ customers instead of $n$, at the price of repeating the computation for every member and for every choice of the part of the source outside $U$.

Fomin et al.\ \cite{fomin} construct set-inclusion families, in which every set of a given size contains a member of a given smaller size. Taking complements turns such a family into a covering family. Their deterministic construction has size and running time within a factor $2^{o(n)}$ of $\binom nt/\binom rt$. We use instead the greedy set-cover rule \cite{chvatal}, whose overhead is polynomial, and count its work in the range required here.

\begin{lemma}[Covering family]\label{lem:family}
For $1\le t\le r\le n$, let $\beta=\binom nt$, $\delta=\binom rt$, $\mu=\binom nr$, and $\rho=\beta/\delta$. A $(t,r)$-covering family with at most $\lceil\rho(1+\ln\beta)\rceil$ members can be constructed deterministically in $\Os(2^n+\mu\rho+\mu\delta)$ time and $\Os(2^n)$ space.
\end{lemma}
\begin{proof}
\emph{Size.} The greedy rule repeatedly selects an $r$-element set that contains the largest number of $t$-element sets not yet contained in a selected set. We call such $t$-element sets \emph{uncovered}. Counting incident pairs shows that each $t$-element set is contained in $\binom{n-t}{r-t}=\mu\delta/\beta$ of the $r$-element sets. If $u$ of the $t$-element sets are uncovered, the numbers of uncovered sets contained in the $\mu$ candidate sets sum to $u\mu\delta/\beta$, so the selected set contains at least $u\delta/\beta=u/\rho$ of them. After $k$ selections at most $\beta(1-1/\rho)^k\le\beta e^{-k/\rho}$ sets remain uncovered. This is at most one after $\lceil\rho\ln\beta\rceil$ selections, and at most one further selection is needed. Since $\rho\ge1$, the total is at most $\lceil\rho(1+\ln\beta)\rceil$.

\emph{Implementation.} Keep one counter per $r$-element set, holding the number of uncovered $t$-element sets it contains and initially equal to $\delta$, and one flag per $t$-element set. Both are indexed by customer masks. Each selection scans the $\mu$ counters, breaking ties by a fixed order, and visits the $\delta$ subsets of size $t$ of the selected set. Over the $\Os(\rho)$ selections this takes $\Os(\mu\rho)$ and $\Os(\rho\delta)=\Os(\beta)$ time. For each newly covered $t$-element set, enumerate the $\mu\delta/\beta$ sets of size $r$ containing it and decrement their counters. Each $t$-element set becomes covered once, so there are at most $\mu\delta$ decrements in all. Enumerating subsets or supersets of a given size takes polynomial time per generated set, and no incidence lists are stored.\end{proof}

\begin{remark}\label{rem:choice}
The choice $r=3t$ balances the number of members against the cost of a single local computation. For any $t\le r\le n$, the proof of Proposition~\ref{pr:restrict} below gives an evaluation bound equal, up to polynomial factors, to
\[
   \frac{\binom nt}{\binom rt}\,2^{n-r}
   \left(2^r + 2^{r(1/2+H((1-t/r)/2))}\right).
\]
In the middle range the second term determines the choice of $r$. With $t$ fixed and $y=t/r$, the part of the base-two logarithm of its entropy upper bound that depends on $r$ is $r[-1/2+H((1-y)/2)-H(y)]$. Its derivative with respect to $r$ is $\tfrac12\log_2(2(1-y)/(1+y))$, which changes sign from negative to positive at $y=1/3$, that is, at $r=3t$. At this choice the second term exceeds the first, because $1/2+H(1/3)>1$.
\end{remark}

\begin{proposition}[Covering-family bound]\label{pr:restrict}
Let $2/7\le x=t/n\le1/3$ and $r=3t$. Given a $(t,r)$-covering family from Lemma~\ref{lem:family}, the layer vector \eqref{eq:layer} can be computed, with witnesses, for every source table $f$ and every $t\le s\le n$, in
\[
             \Os\!\left(2^{\,n(1+H(x)-3x/2)}\right)
\]
time and $\Os(2^n)$ space. Moreover, the construction terms $\mu\rho$ and $\mu\delta$ in Lemma~\ref{lem:family} are bounded by $\Os(10^{n/3})$ and $\Os((7/2^{10/7})^n)$, respectively.
\end{proposition}
\begin{proof}
Let $\mathcal U$ be the covering family.

\emph{Procedure.} For each member $U\in\mathcal U$ and each $J\subseteq V\setminus U$, consider the candidates $(C,S)$ with $S\setminus U=C\setminus U=J$. Write $S=J\cup S'$ and $C=J\cup C'$ with $S',C'\subseteq U$. These candidates form a layer on the customer set $U$, with local source table $f_J(C')=f(J\cup C')$, local target size $s-|J|$, and route size $t$. The capacity test $q(S)-q(C)\le Q$ becomes $q(S')-q(C')\le Q$. The added route serves $S'\setminus C'\subseteq U$, so its cost $\route(S'\setminus C')$ involves only arcs among $U\cup\{0\}$. Skip the pairs $(U,J)$ with $s-|J|\notin[t,r]$. For the other pairs, compute the local layer vector by Proposition~\ref{pr:long} applied to the instance restricted to $U$, and use each local value to update the entry for $S=J\cup S'$.

\emph{Correctness.} Every finite local value is the value of a candidate of the original layer. Conversely, for any candidate $(C,S)$, the $t$-element set $S\setminus C$ lies in some $U\in\mathcal U$. With $J=S\setminus U$ we also have $C\setminus U=J$, because $S\setminus C\subseteq U$. So the candidate occurs in the local computation for $(U,J)$. Thus the minimum of all local values at each target $S$ equals $G^f_{s,t}(S)$, and a local witness $S'\setminus C'=S\setminus C$ is a witness for the original layer.

\emph{Running time and space.} Proposition~\ref{pr:long} on $U$ costs $\Os(2^r+\sqrt{2^r\binom{r}{s'-t}\binom{r}{s'}})$ with $s'=s-|J|$. By \eqref{eq:jensen} with $n$ replaced by $r$ and $x$ by $t/r=1/3$, this is $\Os(2^{r(1/2+H(1/3))})$, since $1/2+H(1/3)>1$. This includes initializing the local tables and transferring the results. There are $\Os(\rho)$ members by Lemma~\ref{lem:family}, and at most $2^{n-r}$ sets $J$ per member. By \eqref{eq:entropy}, $\rho\le(r+1)2^{nH(x)-rH(1/3)}$. Multiplying the three factors gives time $\Os(2^{n(1+H(x))-r/2})=\Os(2^{n(1+H(x)-3x/2)})$. Processing one pair $(U,J)$ at a time and storing the at most $\beta\le2^n$ members of $\mathcal U$ uses $\Os(2^n)$ space.

\emph{Construction terms.} We have $\rho=\prod_{i=0}^{t-1}(n-i)/(r-i)$. The factors are nondecreasing in $i$, and the last one satisfies $(n-t+1)/(2t+1)\le(n-t)/(2t)$ because $3t\le n$. Hence $\rho\le((n-t)/(2t))^t=((1-x)/(2x))^t\le(5/4)^{n/3}$, using $x\ge2/7$ and $t\le n/3$. Thus $\mu\rho\le 2^n(5/4)^{n/3}=10^{n/3}$. Next, $\mu\delta=\binom nr\binom rt$ is the multinomial coefficient with cell counts $(t,2t,n-3t)$, so $\log_2(\mu\delta)/n\le H(3x)+3xH(1/3)$. The derivative of the right side, $3\log_2((1-3x)/(3x))+3H(1/3)$, is decreasing in $x$ and equals $-5$ at $x=2/7$. So the right side is maximized on $[2/7,1/3]$ at $x=2/7$, where the cell proportions are $(2/7,4/7,1/7)$ and their entropy is $\log_2 7-10/7$.\end{proof}

\subsection{The algorithm and the proofs of Theorems~\ref{th:layercover} and~\ref{th:conv}}
Let $\tau=269/930\approx0.289247$, which satisfies $2/7<\tau<1/3$. A layer vector is computed by the split evaluation if $t\le\tau n$, by the covering-family procedure if $\tau n<t<n/3$, and by the ordered-source procedure if $t\ge n/3$. Algorithm~\ref{alg:main} applies this rule within the layered recurrence of Lemma~\ref{lem:recur}. For simplicity it uses the recurrence with budgets also when the fleet is unlimited.

\begin{algorithm}[t]
\caption{Covering the customers with at most $K$ identical vehicles ($K=\infty$ allowed)}\label{alg:main}
\begin{algorithmic}[1]
\State Compute $q$, $P$, $R$, $\route$, and $T$ for all customer sets.
\State Construct a $(t,3t)$-covering family for every $t$ with $\tau n<t<n/3$ by Lemma~\ref{lem:family}.
\State $f(\varnothing)\gets 0$ and $f(S)\gets+\infty$ for $S\ne\varnothing$. \Comment{$f=F_0$}
\For{$b=1,\ldots,\min(K,n)$}
  \State Set $u(\varnothing)\gets0$ and $u(S)\gets T(S)$ for nonempty $S$. Record $S$ as its own last route whenever $T(S)$ is finite.
  \For{$s=1,\ldots,n$ and $t=1,\ldots,\lfloor s/2\rfloor$}
    \If{$t\le\tau n$}
      \State Choose $d$ minimizing $X_d+Y_d$ in \eqref{eq:xy}. Compute $G^f_{s,t}$ by Lemma~\ref{lem:split}.
    \ElsIf{$3t<n$}
      \State Compute $G^f_{s,t}$ by Proposition~\ref{pr:restrict}, using the covering family for $t$.
    \Else
      \State Sort the sources by load. Compute $G^f_{s,t}$ by Proposition~\ref{pr:long}.
    \EndIf
    \State Update $u(S)\gets\min\{u(S),G^f_{s,t}(S)\}$ for all $|S|=s$, recording a witness.
  \EndFor
  \State Save the recorded witnesses for budget $b$ and set $f\gets u$. \Comment{$f=F_b$}
\EndFor
\State Return $f(V)$ and, if it is finite, recover the routes by following the recorded witnesses.
\end{algorithmic}
\end{algorithm}

\begin{proof}[Proof of Theorem~\ref{th:conv}]
There are $O(n^2)$ layer vectors, so it suffices to bound the cost of one of them uniformly in $1\le t\le s\le n$ and in the source table. Let $x=t/n$.
\begin{itemize}
\item If $x\le\tau$, Proposition~\ref{pr:short} with $\alpha=\tau$ and $\kappa=0$ gives the base $\theta_1=2^{1+H(\tau)/2}$.
\item If $\tau<x<1/3$, Proposition~\ref{pr:restrict} applies because $\tau>2/7$. Its exponent $1+H(x)-3x/2$ is decreasing on $[2/7,1/3]$, since $H'(x)=\log_2((1-x)/x)\le\log_2(5/2)<3/2$ there. Its base is therefore at most $\theta_2=2^{1+H(\tau)-3\tau/2}$. The construction in Lemma~\ref{lem:family} adds $\Os(2^n+10^{n/3}+(7/2^{10/7})^n)$.
\item If $x\ge1/3$, Proposition~\ref{pr:long} and \eqref{eq:jensen}, whose right side decreases in $x$, give the base $\theta_3=2^{1/2+H(1/3)}$.
\end{itemize}

\emph{Certificate.} To certify the decimal bound, let $\lambda=67543/25000=2.70172$. Since $2^{H(\tau)}=\tau^{-\tau}(1-\tau)^{-(1-\tau)}$ with $\tau=269/930$ and $1-\tau=661/930$,
\[
 \theta_1^{1860}=\frac{2^{1860}\cdot930^{930}}{269^{269}\cdot661^{661}},\qquad
 \theta_2^{1860}=\frac{2^{1053}\cdot930^{1860}}{\left(269^{269}\cdot661^{661}\right)^2}.
\]
Both are rational, so $\theta_1^{1860}<\lambda^{1860}$ and $\theta_2^{1860}<\lambda^{1860}$ are comparisons of integers after clearing denominators. The remaining comparisons are $\theta_3^6=729/2<\lambda^6$ and, for the construction terms, $10<\lambda^3$ and $7^7/2^{10}<\lambda^7$. The accompanying verification script checks all five inequalities in exact integer arithmetic. Numerically, $\theta_1\approx2.7017112$, $\theta_2\approx2.7017159$, $\theta_3\approx2.6726962$, $10^{1/3}\approx2.1544$, and $7/2^{10/7}\approx2.6005$.

\emph{Conclusion.} Every layer takes $\Os(2.70172^n)$ time. The path tables, covering families, and temporary tables use $\Os(2^n)$ space, and each of the three procedures returns witnesses. By Section~\ref{sec:model} and the assumption on $f$, all integers involved have polynomially many bits, which establishes the stated bounds.\end{proof}

\begin{proof}[Proof of Theorem~\ref{th:layercover}]
\emph{Optimal value.} By Lemma~\ref{lem:recur}, Algorithm~\ref{alg:main} computes the optimal value $F_{K'}(V)$ from $O(n^3)$ layer vectors, each within the bounds of Theorem~\ref{th:conv}.

\emph{Optimal routes.} Whenever an update improves an entry of the table $u$, the algorithm stores the witness, that is, the customer set of the added route. Let $A$ be the witness stored for a finite entry $F_b(S)$ with $A\ne S$, and let $C=S\setminus A$. The stored value is $F_{b-1}(C)$ plus the cost of some route serving $A$, so $F_b(S)\ge F_{b-1}(C)+\route(A)$. The reverse inequality holds because $(C,S)$ is a candidate in \eqref{eq:layer}. Hence a cheapest route serving $A$, recovered from \eqref{eq:hk} by following minimizing predecessors, extends an optimal cover of $C$ with at most $b-1$ routes to an optimal cover of $S$. Repeating from $C$ with budget $b-1$, or stopping when $A=S$, recovers an optimal solution. Storing the witnesses for all budgets costs at most $O(n2^n)$ additional entries.\end{proof}

\subsection{Where the covering families improve the base}\label{sec:crossing}
We now locate the layer that controls the two bounds of Sections~\ref{sec:batch} and~\ref{sec:split} when they are applied to the full customer set. This explains why the covering families are needed (Figure~\ref{fig:bases}).

\emph{The crossing point.} Let $x_*$ be the unique solution in $(0,1/3)$ of
\[
       1+H(x)/2=1/2+H((1-x)/2).
\]
The left side increases and the right side decreases. At zero their values are $1$ and $3/2$, and at $1/3$ the left side exceeds the right side by $(1-H(1/3))/2>0$. Numerically $x_*\approx0.29381537$, and the common base is approximately $2.70719625$.

\emph{Both bounds are attained there.} Consider even $n$ and even $t$ with $t/n\to x_*$, and put $s=(n+t)/2$. This is an admissible layer of \eqref{eq:outer}, since $t\le s/2$ once $3t\le n$. At this layer $X_d=Y_{t-d}$, and
\[
              X_{d+1}/X_d=(t-d)/(n-s+d+1)\le1.
\]
Hence $\max\{X_d,Y_d\}$ is minimized at $d=t/2$, where its value is $\binom{n}{n/2}\binom{n/2}{t/2}$. Stirling's formula gives the limiting base $2^{1+H(x_*)/2}$. For the ordered-source procedure, $\binom{n}{s-t}=\binom ns$ at the same layer, and the expression in \eqref{eq:long} has the limiting base $2^{1/2+H((1-x_*)/2)}$. So the counts in \eqref{eq:xy} and \eqref{eq:long} have the common exponential rate at this layer, even if the cheaper of the two procedures is used at each layer. This is a property of these two counts, and Proposition~\ref{pr:restrict} improves on them in precisely this intermediate range.

\emph{The new crossing point.} The exponent of Proposition~\ref{pr:restrict} meets the exponent of Proposition~\ref{pr:short} where $H(x)=3x$. There is a unique positive root, since the derivative of $H(x)/x$ is $\log_2(1-x)/x^2<0$. The crossing occurs at $x\approx0.28925026$ and gives the base $2^{1+3x/2}\approx2.70171482$. The nearby rational $\tau<0.28925026$ in Algorithm~\ref{alg:main} yields the certified bound above.

\emph{A lower bound at the rational threshold.} The split evaluation in the stated pseudocode enumerates all sets at its selected split size, even in a layer with no finite source. The Python shortcut in Section~\ref{sec:verify} can skip this enumeration. This shows that, for Algorithm~\ref{alg:main}, the base in Theorem~\ref{th:layercover} cannot be lowered below $\theta_1\approx2.7017112$.

\begin{proposition}[Running-time lower bound]\label{pr:tight}
Let $\theta_1=2^{1+H(\tau)/2}\approx2.7017112$. There is a constant $c>0$ such that, on every instance with $n\ge1$ customers and $K\ge1$, the unpruned loops of Algorithm~\ref{alg:main}, when executed on that instance, perform at least $c\,\theta_1^n/n^2$ operations.
\end{proposition}
\begin{proof}
It suffices to prove the bound for all sufficiently large $n$, since $c$ can be decreased to cover finitely many values of $n$.

\emph{A layer that is always computed.} Let $t$ be the largest integer with $t\le\tau n$ such that $n+t$ is even, so that $t>\tau n-2$, and put $s=(n+t)/2$. Since $3t\le n$, we have $t\le s/2$. In its first iteration, Algorithm~\ref{alg:main} therefore computes the layer vector $G^f_{s,t}$ by the split evaluation, with some split size $d$.

\emph{The work does not depend on the data.} For $0<d<t$, the procedure of Lemma~\ref{lem:split} enumerates all $X_d$ first parts and all $Y_d$ last parts, whatever the source table. For $d\in\{0,t\}$ it enumerates $X_0=Y_t$ pairs, and $X_0\ge Y_0$ and $Y_t\ge X_t$. In every case it performs at least $\max\{X_d,Y_d\}$ operations.

\emph{A lower bound on the counts.} With $s=(n+t)/2$, the cell counts of $X_d$ and of $Y_{t-d}$ in the proof of Proposition~\ref{pr:short} coincide, so $X_d=Y_{t-d}$. Moreover $X_{d+1}/X_d=(t-d)/(n-s+d+1)\le1$, because $n-s=(n-t)/2\ge t$. Let $d_0=\lceil t/2\rceil$. If $d\le d_0$, then $X_d\ge X_{d_0}$. If $d>d_0$, then $t-d\le d_0$ and $Y_d=X_{t-d}\ge X_{d_0}$. Hence $\max\{X_d,Y_d\}\ge X_{d_0}$ for every $d$. Since $n$ and $t$ have the same parity,
\[
 X_{d_0}=\binom{n}{\lceil n/2\rceil}\binom{\lfloor n/2\rfloor}{\lfloor t/2\rfloor}.
\]
The first factor is the largest of the $n+1$ binomial coefficients $\binom nk$, which sum to $2^n$, so it is at least $2^n/(n+1)$. The ratio $\lfloor t/2\rfloor/\lfloor n/2\rfloor$ equals $t/n$ or $(t-1)/(n-1)$, and in both cases it lies between $\tau-3/n$ and $\tau$. For large $n$ this range lies in $[1/5,1/2]$, where $H$ is increasing with derivative at most $2$. By \eqref{eq:entropy}, the second factor is therefore at least $2^{\lfloor n/2\rfloor(H(\tau)-6/n)}/(n+1)\ge2^{(n/2)H(\tau)-4}/(n+1)$. Multiplying the two factors gives $X_{d_0}\ge\theta_1^n/(16(n+1)^2)$.\end{proof}

Together, Theorem~\ref{th:layercover} and Proposition~\ref{pr:tight} place the worst-case exponential base of Algorithm~\ref{alg:main} with $\tau=269/930$ between $2.70171$ and $2.70172$. For the variant switching at the exact crossing $x_0$ of $H(x)=3x$, the two bounds match. Indeed, the preceding lower-bound proof applies with $\tau=x_0$, and the short-route and covering-family upper exponents then both equal $1+3x_0/2$. The other bases in the proof of Theorem~\ref{th:conv} are strictly smaller. Thus this variant has base $2^{1+3x_0/2}\approx2.70171482$ up to polynomial factors. The switch $t\le x_0n$ is exactly the integer test $8^tt^t(n-t)^{n-t}\le n^n$, with $0^0=1$ at $t=n$.

The bound is not special to the first iteration, in which the source table is $F_0$, because the enumeration in the split evaluation is the same for every source table. The reference implementation of Section~\ref{sec:verify} performs the same enumeration in the split evaluation. It is faster only on layers to which its shortcut applies or for which no route of size $t$ is feasible.

\section{Consequences and extensions}\label{sec:fleet}
Theorem~\ref{th:conv} places no restriction on the first operand or on the route size. It therefore applies when a particular vehicle must be processed next, so that the added route need not be a smallest one and the layer in \eqref{eq:layer} can have $t>s/2$. This accounts for a difference between the two analyses. The split-only bound of Proposition~\ref{pr:splitonly} uses the freedom to remove a smallest route, whereas Theorem~\ref{th:conv} holds for all $1\le t\le s$. Sections~\ref{sec:listed} and~\ref{sec:variants} are consequences of Theorem~\ref{th:conv} and of Algorithm~\ref{alg:main}. Sections~\ref{sec:windows} and~\ref{sec:resources} modify the procedures.

\subsection{An explicitly listed heterogeneous fleet}\label{sec:listed}
\begin{corollary}\label{co:fleet}
Suppose the input lists $m$ vehicles. Vehicle $k$ may serve at most one route, has capacity $Q_k$, and has its own nonnegative integer directed costs, including departure costs from a prescribed start terminal and return costs to a prescribed end terminal. Vehicles may be left unused. With common nonnegative integer customer demands, a minimum-cost feasible set of routes can be found in $\Os(2.70172^n)$ time and $\Os(2^n)$ space.
\end{corollary}
\begin{proof}
Let $F^{(k)}(S)$ be the minimum cost of serving $S$ with a subset of the first $k$ vehicles, each serving at most one route. Set $F^{(k)}(\varnothing)=0$ for every $k$ and $F^{(0)}(S)=\INF$ for nonempty $S$. This also settles the case of an empty vehicle list. For vehicle $k$, build $P_k$, $R_k$, $\route_k$, and $T_k$ from its own costs and capacity, using its departure costs in the forward base case and its return costs in the reverse base case. For nonempty $S$,
\begin{equation}\label{eq:fleetrec}
 F^{(k)}(S)=\min\left\{F^{(k-1)}(S),\ (F^{(k-1)}\star T_k)(S)\right\}.
\end{equation}
The first term leaves vehicle $k$ unused. Every solution that uses vehicle $k$ assigns it exactly one nonempty customer set and is represented in the second term, by \eqref{eq:conv}. Theorem~\ref{th:conv} applies to each of the $m$ convolutions, because its bounds are uniform in the source table, in the capacity, and in the arc costs. The number of vehicles $m$ is polynomial in the length of an explicitly listed fleet. For each finite entry, store either an idle-vehicle flag or the minimizing route set. Follow these choices backwards, and recompute individual path tables during the recovery of the routes if necessary. These costs are absorbed in the stated bounds.\end{proof}

If the fleet is given by vehicle types with multiplicities, a type with $K_j$ identical vehicles is processed by $\min\{K_j,n\}$ successive applications of the recurrence \eqref{eq:fleetrec} with the same kernel. The multiplicities may therefore be encoded in binary.

\subsection{Further variants}\label{sec:variants}
The next corollary collects variants that need no new procedure. Open routes \cite{sariklis}, optional customers that may be outsourced at a penalty \cite{bolduc}, and lower bounds on the number of customers of a route \cite{gouveia} are established variants in the routing literature.

\begin{corollary}\label{co:variants}
For identical vehicles, the bounds of Theorem~\ref{th:layercover} also hold for the following variants.
\begin{enumerate}
\item[(a)] \emph{Fleet-size objectives.} Computing, for every $b\le\min\{K,n\}$, the minimum cost of a solution with at most $b$ routes. This gives the minimum number of vehicles and a cheapest solution with that number, and it covers a fixed cost per vehicle.
\item[(b)] \emph{Open routes.} Routes that end at their last customer, or at an end depot different from the start depot.
\item[(c)] \emph{Optional customers.} Each customer $i$ may be left unserved at a nonnegative integer penalty $\pi_i$.
\item[(d)] \emph{Cardinality bounds.} Each route serves at least $t_{\min}$ and at most $t_{\max}$ customers, where $1\le t_{\min}\le t_{\max}$.
\end{enumerate}
Variants (b), (c), and (d), and vehicle-specific fixed costs, are also covered by Corollary~\ref{co:fleet}.
\end{corollary}
\begin{proof}
Variants (a), (b), and (c) use Algorithm~\ref{alg:main} as it stands, and (d) uses Theorem~\ref{th:conv}.

(a) After iteration $b$, Algorithm~\ref{alg:main} holds the table $F_b$, so the values $F_b(V)$ are by-products. A fixed cost per vehicle is added to every $c_{0i}$.

(b) No argument uses a relation between the costs $c_{0i}$ and $c_{i0}$. Set $c_{i0}=0$, or let $c_{i0}$ be the cost from $i$ to the end depot.

(c) Algorithm~\ref{alg:main} computes $F_{K'}(S)$ for every $S\subseteq V$. Return the minimum of $F_{K'}(S)+\sum_{i\notin S}\pi_i$ over all $S$, which takes $O(2^n)$ further operations.

(d) Call a route feasible if it also satisfies the cardinality bounds, and define covers and $F_b$ accordingly. Removing any one route from a cover gives, as in \eqref{eq:conv}, that $F_b(S)$ is the minimum of $G^{F_{b-1}}_{s,t}(S)$ over $t_{\min}\le t\le\min\{t_{\max},s\}$. Theorem~\ref{th:conv} covers each of these layers, including those with $t>s/2$.

The same changes apply to each vehicle of a listed fleet.\end{proof}

\subsection{Load windows}\label{sec:windows}
A load window constrains total demand on each route. With unit demands, a lower load bound is a lower bound on the number of customers, as studied by Gouveia et al.\ \cite{gouveia}. Suppose that every route must have a load of at least $Q_{\min}$ and at most $Q$, where $0\le Q_{\min}\le Q$. Then $T(A)=\route(A)$ if $Q_{\min}\le q(A)\le Q$ and $T(A)=\INF$ otherwise, and the minimum in \eqref{eq:layer} carries the additional constraint $q(C)\le q(S)-Q_{\min}$. The sources admissible for a target then form an interval of the sorted list instead of a prefix. The next lemma extends Lemma~\ref{lem:order} to this situation.

\begin{lemma}[Ordered sources with intervals]\label{lem:interval}
In the setting of Lemma~\ref{lem:order}, suppose that each target $S$ accepts the sources at the positions $a(S)\le i\le k(S)$ of the list, where $a(S)$ and $k(S)$ can be computed in polynomial time, and that assumption (ii) holds for every set of consecutive sources $C_a,\ldots,C_b$ in place of a prefix. Then the minima $\min_{a(S)\le i\le k(S)}w(C_i,S)$ for all $S\in\targets$ can be computed in $\Os((\Gamma+\sqrt{\Gamma NM})\log(2N))$ time and $\Os(\Gamma)$ space, with minimizing sources if the procedures in (i) and (ii) return them.
\end{lemma}
\begin{proof}
Choose $g$ and $h=\lceil N/g\rceil$ as in the proof of Lemma~\ref{lem:order}, and divide the positions $1,\ldots,N$ into consecutive blocks of width $h$, the last block possibly shorter. There are $\lceil N/h\rceil\le g$ blocks.

\emph{Decomposition of an interval.} For a target $S$, the blocks contained in its interval are consecutive, and fewer than $2h$ positions of the interval lie outside them. Evaluate the scores at these positions directly. If $\nu\ge1$ blocks are contained in the interval, let $p$ be the largest power of two with $p\le\nu$. Since $2p>\nu$, the first $p$ and the last $p$ of these blocks together contain all of them.

\emph{Runs.} Call a set of $p$ consecutive blocks, with $p$ a power of two, a \emph{run}. There are at most $g(1+\log_2g)$ runs, and each target uses at most two of them. For each run that is used by some target, call (ii) once on the consecutive sources of the run, and use the result for all targets that use the run. The required minimum for $S$ is the smallest of the directly evaluated scores and of the at most two run minima. The two runs may overlap, which does no harm because a minimum is taken.

\emph{Running time and space.} The time is $\Os(g\Gamma\log(2g)+Mh)$. With the choice of $g$ from the proof of Lemma~\ref{lem:order}, this is $\Os((\Gamma+\sqrt{\Gamma NM})\log(2N))$. Processing the runs one at a time and retaining, for each target, the best value found so far and a minimizing source uses $\Os(\Gamma)$ space.\end{proof}

\begin{proposition}[Load windows]\label{pr:window}
With load windows, Theorems~\ref{th:layercover} and~\ref{th:conv} and Corollaries~\ref{co:fleet} and~\ref{co:variants} remain valid, with vehicle-specific windows in Corollary~\ref{co:fleet}.
\end{proposition}
\begin{proof}
Lemmas~\ref{lem:recur}, \ref{lem:multi}, and~\ref{lem:family}, the reduction in Proposition~\ref{pr:restrict}, and the arguments of Sections~\ref{sec:listed} and~\ref{sec:variants} do not depend on the form of the load test, and the counting bounds are unchanged. Two procedures change.

\emph{Ordered-source procedure.} In Proposition~\ref{pr:long}, the sources admissible for a target $S$ are those with $q(S)-Q\le q(C)\le q(S)-Q_{\min}$. In the list sorted by load they occupy consecutive positions, whose ends are found by binary search. Lemma~\ref{lem:multi} applies to any family $\mathcal P$, in particular to consecutive sources. Every source of a run used by $S$ is admissible for $S$, so the merging in Lemma~\ref{lem:multi} remains valid. Lemma~\ref{lem:interval} therefore gives the bound of Proposition~\ref{pr:long} up to the factor $\log(2N)=O(n)$.

\emph{Split evaluation.} In Lemma~\ref{lem:split}, the entries of $\mathcal L_i$ admissible for a last part $B$ are those with $Q_{\min}-q(B)\le q(A)\le Q-q(B)$. They are consecutive in the sorted list. Replace the running minima by a table of the minima over all sets of $2^k$ consecutive entries, which is built with $O(\log|\mathcal L_i|)$ operations per entry and answers each query from two overlapping sets. For $d\in\{0,t\}$ the window is tested directly. The proof of Lemma~\ref{lem:split} then applies with the two-sided test in place of the one-sided one.\end{proof}

\subsection{Several capacity constraints}\label{sec:resources}
Give each customer $i$ a nonnegative integer demand $q_i^e$ and each vehicle a capacity $Q^e$ in each coordinate $e=1,\ldots,D$. A route must satisfy all $D$ upper bounds. For example, weight and volume give $D=2$. In the layer for target $S$, each coordinate imposes a threshold $q^e(C)\ge q^e(S)-Q^e$ on the source $C$. Their intersection is an orthant query on the source load vectors.

\begin{proposition}[Several resources]\label{pr:multi}
For any fixed $D\ge1$, all layer vectors and their witnesses in Theorem~\ref{th:conv} can be computed under $D$ capacity constraints in deterministic $\Os(2.70172^n)$ time and $\Os(2^n)$ space. The same bounds hold for identical vehicles with or without a fleet limit and for an explicitly listed fleet with vehicle-specific capacity vectors, costs, and terminals.
\end{proposition}
\begin{proof}
\emph{Long routes.} Fix $s,t$. Let $N$ be the number of finite sources of size $s-t$, $M=\binom ns$ the number of targets, and $\Gamma=2^n$. If $NM\le\Gamma$, test every source--target pair directly. Otherwise, sort the $N$ sources by nonincreasing load separately in each coordinate, breaking ties by mask. In order $e$, binary search finds the prefix of sources with $q^e(C)\ge q^e(S)-Q^e$, including all ties at the threshold. The rank vectors of admissible loads lie in the product of these prefixes. Containment $C\subset S$ will be enforced when evaluating the route cost.

In each coordinate, put the ranks at the leaves of a complete binary tree, padding to the next power of two. Write $L=1+\lceil\log_2 N\rceil$. Each rank belongs to at most $L$ node intervals, and each prefix is a disjoint union of at most $L$ such intervals. A product of $D$ intervals is a box. Each source belongs to at most $L^D$ boxes, and each target query decomposes into at most $L^D$ disjoint boxes. Enumerate the ancestor tuples of each source and sort by tuple to build the lists for all nonempty boxes, with at most $NL^D$ memberships in total. Target lists can likewise be grouped deterministically by box.

Take $h=\lceil\sqrt{N\Gamma/M}\rceil$. For each queried box with at most $h$ sources, check containment and evaluate the individual scores using the precomputed tour table. This takes $O(ML^Dh)$ score evaluations. There are at most $NL^D/h$ larger boxes. For each queried large box, apply Lemma~\ref{lem:multi} once to its source list and use the resulting minima only at targets querying that box. Every source in the box satisfies all $D$ load tests for each such target, and the path recurrence enforces containment. Taking the minimum over the query boxes therefore returns the layer value and a minimizing source for every finite entry.

Processing large boxes successively requires $\Os(\Gamma+(N+M)L^D)$ space. The time, up to polynomial factors, is
\[
 (N+M+Mh+\Gamma N/h)L^D.
\]
Here $N,M\le\Gamma$ and $L\le n+1$. For fixed $D$, rounding $h$ gives the long-route bound $\Os(\Gamma+\sqrt{\Gamma NM})$ and space $\Os(\Gamma)$, as in Proposition~\ref{pr:long}.

\emph{Short and intermediate routes.} For each list $\mathcal L_i$ in Lemma~\ref{lem:split}, sort the first parts $A$ by \emph{increasing} load in each coordinate. Store a minimum and a minimizing first part in each nonempty box. For a last part $B$, query the prefixes defined by $q^e(A)\le Q^e-q^e(B)$ for all $e$ and take the smallest box minimum. This is an orthogonal range-minimum query \cite{bentley}. The work per entry or query grows by at most a polynomial in $n$ for fixed $D$, so the counts $X_d,Y_d$ still give the short-route bound. Empty lists give $\INF$, and the endpoint splits test all capacities directly.
For intermediate routes, fix a covering member $U$ and the outside set $J$ as in Proposition~\ref{pr:restrict}. In every coordinate $q^e(J\cup S')-q^e(J\cup C')=q^e(S')-q^e(C')$ for $S',C'\subseteq U$. The local long-route calculation thus has the same time bound on $|U|$ customers and returns a valid route witness. The proof of Theorem~\ref{th:conv} now gives the stated bounds for every source table. The fleet recurrences in Lemma~\ref{lem:recur} and Corollary~\ref{co:fleet} give the routing claims.\end{proof}

\section{When customer groups can be visited consecutively}\label{sec:savings}
For distinct customers $i,j$, the saving from joining them directly is $s_{ij}=c_{i0}+c_{0j}-c_{ij}$. This is the directed form of the savings used by Clarke and Wright \cite{clarkewright}. For $n\ge2$ set $s=\min_{i\ne j}s_{ij}$ and $e_{ij}=s_{ij}-s$. All $e_{ij}$ are nonnegative, even when some $s_{ij}$ are negative. If $C_0=\sum_i(c_{0i}+c_{i0})$, the cost of a cover with $b$ routes is exactly
\begin{equation}\label{eq:savingtotal}
 C_0-s(n-b)-\sum_{i\to j}e_{ij},
\end{equation}
where the sum is over consecutive pairs of customers. Fix $z\ge0$. Form an undirected graph by joining $i,j$ whenever $\max(e_{ij},e_{ji})>z$. Every arc between different components has $e_{ij}\le z$. Within a component $G$, form a bipartite graph on a left and a right copy of $G$, with an edge for each ordered pair $i\ne j$ for which $e_{ij}\le(n-1)z$. Let $d_G$ be its maximum matching size. A \emph{run} of $G$ is a maximal consecutive sequence of its customers in a route.

\begin{theorem}\label{th:saving}
Suppose the components can be packed whole into $b_*\le K$ nonempty vehicles of capacity $Q$, where $s(b_*-b)\le0$ for every feasible route count $b$. For $z=0$, some optimum visits every component in a single run. For $z>0$, every optimum visits each component $G$ in at most $\max(1,d_G)$ runs in total over all routes.
\end{theorem}
\begin{proof}
Take any feasible cover and decompose it into component runs. Preserve all arcs within each run. Order the runs of each component, concatenate them, and assign the whole components to the specified packing. The lost intercomponent arcs have total $e$ at most $(n-1)z$ and every new arc has $e\ge0$. The route-count term in \eqref{eq:savingtotal} can only improve. This proves the assertion for $z=0$.

If $G$ has $a\ge2$ runs, place them in a cyclic order. The $a$ proposed joins from one run's end to the next run's beginning have distinct tails and distinct heads. If $a>d_G$, one join has $e>(n-1)z$. Break the cyclic order at a different join. Putting these runs together now recovers more than the total lost over all components, giving a strictly cheaper feasible solution.\end{proof}

With one vehicle the packing condition holds whenever the instance is feasible. For several vehicles, an explicit sufficient test when $s>0$ is a packing into $\max(1,\lceil q(V)/Q\rceil)$ vehicles for $Q>0$. For $s=0$, any feasible whole-component packing suffices. For $s<0$, a packing into $\min(K,n)$ nonempty vehicles suffices. A subset dynamic program on the components tests a proposed upper limit by minimizing, in order, the number of opened bins and the load in the last bin. For the last case a packing with fewer bins can be split if it contains at least as many components as the desired number of bins.

For example, let $Q=10$, $K=3$, and let the four demands be $(3,3,6,6)$. Set every off-diagonal arc cost to 10 except for the two arcs between the customers of demand 3, each of which costs 9. At $z=0$ these two customers form one component and the other two are singletons. The minimum saving is $s=10$. Two routes pairing each demand 3 with a demand 6 cost 60 in total. This is optimal: every cover with two routes must use these pairings, and any cover with three routes costs at least 69. Keeping the component whole requires three routes and attains cost 69. Its feasible packing has $s(3-2)>0$, which is precisely the case excluded by Theorem~\ref{th:saving}.

\begin{corollary}\label{co:blocks}
Contract all components certified to be single runs, retaining every other customer as a singleton. If the resulting partition has $m$ blocks and the largest block has $k$ customers, the optimum and routes take $\Os(2^{\max(m,k)})$ time. At $z=0$ with one vehicle, $\Os(2^k)$ suffices.
\end{corollary}
\begin{proof}
For each block $J$ compute the cost $H_J(u,v)$ of a cheapest Hamiltonian path within $J$ for each pair of endpoints, including a singleton path of cost zero. Its subset table takes $\Os(2^{|J|})$. Appending $J$ after physical endpoint $i$ at final endpoint $v$ costs $\min_u(c_{iu}+H_J(u,v))$. Apply the slack algorithm to subsets of blocks, with their total demands and the physical last customer as endpoint. The slack test depends only on the block union. Theorem~\ref{th:saving} puts an optimum in this restricted problem, and storing minimizing choices recovers all visits. For a proposed route count satisfying the condition in Theorem~\ref{th:saving}, the whole-component packing test takes at most $\Os(2^m)$, since the number of components is at most $m$. At $z=0$, solve a separate depot tour on each component, remove its depot endpoints and concatenate the paths according to the packing. Every link between components has the same saving $s$, so each independent optimization gives the correct objective for that packing. For one vehicle the packing is immediate.\end{proof}

Test $z=0$ and the distinct values $\max(e_{ij},e_{ji})$ over unordered customer pairs. The component partition changes only at these values. At a fixed component partition the smallest threshold gives the strongest matching test. A dense component can be certified even if every outgoing internal arc of one customer has a small saving, since these arcs have matching number one.

\section{The boundary at one vehicle}\label{sec:limits}
\emph{A bound for a computational model.}
Set $K=1$ and $Q\ge q(V)$. An optimal solution is a minimum-weight directed Hamiltonian cycle through the depot and all $n$ customers. Jerrum and Snir \cite[Section~4.4]{js} prove that the Hamiltonian circuit function on $n+1$ vertices requires at least $n((n-1)2^{n-2}+1)$ additions in a pure straight-line min-plus circuit. Its wiring is fixed independently of the arc costs and its operations are minima and ordinary addition. For nonnegative integer inputs and nonnegative scalar constants, the bound transfers by scaling rational inputs to integer multiples, dividing the output by the scale, and passing to the limit. This replaces each fixed finite constant by zero and gives a circuit correct on the rationals, hence on the nonnegative reals by continuity. For fixed demands the slack construction uses only minima and additions on cost values. The worst-case exponential base is therefore two within this model. Cost-dependent branching or other arithmetic falls outside the lower bound. Koana and Kumabe's randomized algorithm for unweighted directed Hamiltonicity \cite{kk} concerns a different objective and computational guarantee.

\emph{Three equally spaced anchors.}
Write $N$ for the number of TSP vertices, including the depot. For $N$ divisible by three, choose three ordered anchors and compute the cheapest elementary path between each consecutive pair on each set of $N/3-1$ interior vertices. An incumbent permits a segment set to be discarded when its price plus the minimum prices of the other two segments already reaches the incumbent. One remaining segment fixes a vertex assignment, and two remaining segments give a binary choice. Three tables joined on the binary choices shared by adjacent segments form a minimum-weight triangle. With the deterministic integer min-plus product analyzed by Stoian \cite[Appendix~A]{stoian}, Appendix~\ref{app:anchors} obtains an upper bound of $\Os(1.999972^N+2^r)$, where $r$ is the binary rank of the incidence vectors of the unresolved anchor triples. A tour still improving the incumbent must give an unresolved triple at every simultaneous rotation of its three anchors. Its visited set after any number of these steps is a sum of such incidence vectors. A rank close to $N$ can leave almost all middle-layer masks in the dynamic program.

\begin{proposition}\label{pr:separation}
There is a family of directed metric TSP instances on $N=27L$ vertices, $L\ge12$, on which the rotating three-anchor procedure of Appendix~\ref{app:anchors} visits $2^{N-O(\log N)}$ masks, even when given the optimal tour as incumbent. On the same instances Theorem~\ref{th:saving} gives an exact algorithm in $\Os(2^{2N/3})$ time.
\end{proposition}
\begin{proof}[Proof outline]
Start with an arbitrary directed core on $18L$ vertices and attach $L$ copies of the Petersen metric at the depot, adding nine vertices per copy. A tour of a copy has cost 11 in its metric units. A fractional collection of elementary fragment paths can cover the same nine vertices at cost $81/8$. Scaling the copies above all core costs makes this gap available for every choice of the three anchors, including anchors inside the copies. A middle-layer choice of nearly half the core and nearly half the added vertices is still affordable as a partial tour. Appendix~\ref{app:anchors} gives the metric, the fractional witnesses and the counting argument. In the savings graph, the $18L-1$ core customers form a certified component. Taking every other customer as a singleton gives at most $9L+1$ blocks with maximum size $18L-1$. Corollary~\ref{co:blocks} proves the upper bound.\end{proof}

In this proposition the lower bound counts states visited by the specified procedure. The same inputs admit a faster algorithm, so the example also shows why such a state count cannot establish a complexity lower bound for TSP or CVRP.
\section{Computational verification}\label{sec:verify}
\emph{Base two and the load refinement.} The exact-count implementation checks 132,096 integer cutoff inequalities and agrees with unrestricted submask enumeration on 119,632 table entries, including the witnesses for finite entries. Of 283 full-instance checks, 280 use enumeration of customer permutations and route cuts. Three additional instances include the 15-customer Augerat P-n16-k8, for which the returned cost is 450. Tests include empty inputs, zero demands, zero capacity and costs shifted by 1,200 bits. The load-sensitive solver agreed with its comparator on all 100 instances, including infeasible cases. Of these, 84 were compared with permutation-and-cut enumeration and 16 with the base-two solver. The split was used on 40 instances. Appendix~\ref{app:count} proves the intermediate route-count bound.

\emph{Published benchmark optima.} Table~\ref{tab:bench} compares the returned costs with the optima in the COIN-OR library \cite{coin}. Each returned solution was checked for customer coverage, capacity, fleet size, and total cost recomputed from its arcs. P-n16-k8 also appears in the full-instance checks above. The P-n16-k8 row uses the load-sensitive solver. The other two rows use Algorithm~\ref{alg:slack}. The two Augerat instances use Euclidean distances rounded to the nearest integer on each arc. E-n13-k4 uses the supplied integer distance matrix.

\begin{table}[htbp]
\centering\small
\caption{Objective values on public CVRP instances. The fleet limit is the maximum number of routes allowed in the run.}\label{tab:bench}
\begin{tabular}{@{}lrrrrrr@{}}
\toprule
Instance & Customers & Capacity & Fleet limit & Routes used & Returned & Published\\
\midrule
P-n16-k8 & 15 & 35 & 8 & 8 & 450 & 450\\
P-n19-k2 & 18 & 160 & 2 & 2 & 212 & 212\\
E-n13-k4 & 12 & 6,000 & 4 & 4 & 247 & 247\\
\bottomrule
\end{tabular}
\end{table}

\emph{Structural certificates.} The three-fragment verification includes 640 table joins against enumeration, 93 checks of TSP tours or the restricted rotation problem, and 107 checks of the savings and packing procedures. On a generated directed matrix with 324 vertices, the savings test identified an exponent of 215. The test computes the certificate only. It does not solve that instance by executing the $2^{215}$ dynamic program. Appendix~\ref{app:anchors} gives the construction and the fractional certificates.

\emph{Arbitrary-source implementation.} A reference implementation of the batched algorithm, written in Python with only the standard library, accompanies the paper as supplementary material. It uses integer costs and exact comparisons, and it selects the procedure for each layer as in Algorithm~\ref{alg:main}. It differs from Algorithm~\ref{alg:main} in three ways. With an unlimited fleet it evaluates the recurrence once, in increasing order of $s$, as described after Lemma~\ref{lem:recur}. It skips route sizes for which no route is feasible. And it contains a shortcut: for any $(s,t)$ it enumerates the customer sets of feasible routes of size $t$ directly if their number multiplied by $\binom{n-t}{s-t}$ is at most $n2^n$. This test counts all possible sources for such sets and keeps the shortcut within the worst-case bound. The shortcut is taken for every evaluated layer on P-n16-k8. Its use on other instances depends on the displayed count. The verification therefore tests the three procedures separately, and with the shortcut disabled. For routing tables the implementation uses an integer sentinel greater than $2n\cmax$. When supplying an external table to a layer routine, choose its infinity parameter greater than $\max_{C:f(C)<\infty} f(C)+2n\cmax$. Use that same sentinel for missing source entries and initial output values. If all source entries are infinite, every output is infinite.

\emph{Whole instances.} For each $n=0,\ldots,6$, twelve generated instances were each solved with an unlimited fleet and with a generated fleet limit. The comparator enumerates all customer permutations and all possible ways of cutting a permutation into routes. Two further eight-customer instances have known answers, one with zero loads and capacity and one with an insufficient fleet. The resulting 170 checks compare objective values and verify that the returned routes meet the fleet and capacity constraints, visit every customer once, and sum to the reported objective. The generated cases include directed costs and integers scaled by $2^{1200}$. All layers of these instances are computed by the shortcut, so these checks validate the layered recurrence, the fleet limit, and the recovery of routes, but not the three procedures.

\emph{Single layers.} Direct tests bypass the choice of procedure. On three 12-customer demand profiles and seven $(s,t)$ pairs, the split evaluation and the ordered-source procedure were each compared at every target with direct evaluation of \eqref{eq:layer}. All 17,634 values agreed, and each finite entry supplied a witness that attains the minimum. These checks include equal loads and the split sizes $d=0$ and $d=t$. The covering-family procedure was separately compared on 2,864 targets with $n=10,11,13$, including arbitrary finite source values, infinite entries, and large binary costs. The tests verify that every $t$-element set lies in a member of the family and that every finite local result supplies a witness of the original layer.

\emph{The solver without the shortcut.} With the shortcut disabled, every layer is computed by one of the three procedures. Eleven instances with $n=10,\ldots,13$, with and without a fleet limit, were solved in this way, and the complete final table was compared with the recurrence $F_b=F_{b-1}\star T$ evaluated by enumeration of all subsets. All 33,792 table entries agreed, and every finite entry was explained by its recorded witness. Two of the eleven instances are infeasible for the full customer set. The runs comprise 443 layers computed by the split evaluation, of which 173 use a split size $0<d<t$ and 270 the enumeration for $d\in\{0,t\}$, as well as 73 layers computed by the covering-family procedure and 113 computed by the ordered-source procedure. In the first iteration of a run with a fleet limit the source table is $F_0$, so these layers have no finite source. At least one source is finite in 119, 152, 47, and 78 of the layers of the four kinds, respectively. The covering-family procedure is selected only when some integer $t$ satisfies $\tau n<t<n/3$, which first happens for $n=10$ and $n=13$.

\emph{Larger instances and listed fleets.} The full solver recovered a known zero-cost solution on a directed 14-customer instance, using the split evaluation and the ordered-source procedure for three layers each. It also recovered the optimal cost $450$ on the 15-customer Augerat instance P-n16-k8 with $Q=35$ and at most eight vehicles, as tabulated by COIN-OR \cite{augerat,coin}; on this instance every layer is computed by the shortcut. Two empty-fleet boundary checks give cost zero for an empty customer set and infeasibility for a nonempty set. A separate check of Corollary~\ref{co:fleet}, which calls the three procedures for all $1\le t\le s$ without the shortcut, agreed on 72 small listed-fleet instances with exhaustive enumeration of customer permutations and cuts that allow unused vehicles. On a ten-customer instance with vehicle-specific costs and capacities seven and three, the listed-fleet implementation recovered the prescribed zero-cost routes, using the covering-family procedure for the route of size three.

\emph{Extensions.} A separate script checks the extensions of Section~\ref{sec:fleet}. Its layer procedures are written independently of the solver, and it takes the path tables and the covering families from the solver. With load windows, the interval form of the ordered-source procedure, the split evaluation with range minima, and the covering-family procedure agreed with direct evaluation of the layer vector in 2,952, 2,190, and 815 comparisons of target values, respectively. Here a target is compared once for every block width or split size that is tested. The split evaluation with two capacity constraints agreed in 2,230 comparisons. For two and three resources, the orthant procedure agreed with direct evaluation in 6,390 target comparisons. Applying it inside the covering family gave agreement in another 1,360 comparisons, including cases with a nonempty fixed outside set. The combination step in the proof of Proposition~\ref{pr:splitonly}, applied to tables $F_b$ obtained by direct enumeration, agreed with exhaustive enumeration in 54 cases, of which 17 are infeasible. These cases have one or two capacity constraints and include instances whose capacity forces every route to be small. In the same way, the fleet-size by-products, optional customers, and cardinality bounds of Corollary~\ref{co:variants} agreed with exhaustive enumeration in 44, 8, and 8 cases, of which 10, none, and 2 are infeasible. The script also confirms the identities and the lower bound on the counts in the proof of Proposition~\ref{pr:tight} for every $n$ from $8$ to $400$.

The ancillary files contain the reference implementations, test generators, random seeds, comparators, and integer certificate. The benchmark script reads the original COIN-OR instance files. The counts above describe correctness checks.

\section{Concluding remarks}\label{sec:conclusion}
The unused capacity of a removable route suffices to recover an exact optimum in $\Os(2^n)$ time for the homogeneous, single-capacity model with binary data. For $Q>0$ and individual demands at most $Q$, total demand in vehicle loads determines an explicit saving in the exponent. An arbitrary source table has no removable-route interpretation, and for that operation the structured convolution gives base $2.70172$.

The principal open question is a deterministic $\Os((2-\varepsilon)^n)$ algorithm for a fixed $\varepsilon>0$ on \emph{all} inputs. It includes arbitrary binary weighted directed TSP as a special case. The min-plus circuit lower bound and the explicit-state lower bound of Section~\ref{sec:limits} identify operations or representations that a faster algorithm would have to avoid. Neither establishes a general time lower bound. Polynomial-space exact routing with a magnitude-independent base below three, and comparable bounds under time windows, are further questions.

\subsection*{Use of generative artificial intelligence}
The author used OpenAI Codex to help review the literature, develop the algorithm, and check proofs, and Claude to help draft the manuscript, formulate extensions, and write test scripts. The author conceived the original idea and directed the research. Codex also assisted in synthesizing the extensions and preparing this revision.

\appendix
\input{appendices}

\end{document}

%% file: appendices.tex
\section{The intermediate route-count bound}\label{app:count}
\begin{theorem}
Single-depot capacitated vehicle routing on $n$ customers with identical vehicles,
binary nonnegative integer demands, capacity, and directed arc costs can be solved
deterministically in $\Os((1+2^{3/4})^n)$ time and $\Os(2^n)$ space. The algorithm
returns an optimal set of routes or reports infeasibility, with or without a fleet limit.
\end{theorem}

The exponential base is $2.6817928305\ldots$. The arbitrary-source algorithm of Theorem~\ref{th:conv} gives $2.70172$ for this model. Its worst split layer has $t/n\approx0.28925$ and $s/n\approx0.64463$, where
$s$ counts customers covered and $t$ counts customers on the route added. Here
$s/3<t<s/2$. The recurrence below skips this entire interval.

\subsection{Choosing the last route}
Let $T(A)$ be the minimum cost of a capacity-feasible tour visiting exactly
$A$, with $T(\varnothing)=+\infty$. Held--Karp gives this table, directed path
tables, and tour witnesses in $\Os(2^n)$ time and space. Let
\[
 B(S)=\min_{\varnothing\ne A\subsetneq S}\{T(A)+T(S\setminus A)\}.
\]
This is the cost of using exactly two nonempty routes. Write $F_b(S)$ for the
minimum cost with at most $b$ routes, and $G^f_{s,t}(S)$ for the minimum
of $f(S\setminus A)+T(A)$ over $t$-subsets $A\subseteq S$, where $|S|=s$.
An empty minimum is $+\infty$. Set $F_b(\varnothing)=0$.
For nonempty $S$, $F_1(S)=T(S)$, $F_2(S)=\min\{T(S),B(S)\}$, and for $b\ge3$,
\[
 F_b(S)=\min\!\left\{T(S),\;B(S),\;
          \min_{1\le t\le\lfloor s/3\rfloor}G^{F_{b-1}}_{s,t}(S)\right\}.
 \tag{A.1}\label{eq:app-small-route}
\]
Every finite term is a feasible cover within budget. Conversely, $T$ and $B$
account for optima using one and two routes. In any optimum with at least
three routes, a smallest route serves at most $\lfloor s/3\rfloor$ customers.
Removing it leaves at most $b-1$ routes. This proves \eqref{eq:app-small-route}.

In computing $B$ either route can be added last, so it suffices to evaluate
$G^T_{s,t}$ for $\lceil s/2\rceil\le t<s$. The two-route table is computed
once and shared across all budgets. Sections below bound its cost and the cost
of all small-route extensions in \eqref{eq:app-small-route}.

\subsection{Covering the target and computing the two-route table}
The ordered-source procedure of Proposition~\ref{pr:long} computes a
layer on $r$ customers, for any first operand $f$, in
\[
 \Os\!\left(2^r+\sqrt{2^r\Bin r{s-t}\Bin rs}\right)
 \tag{A.2}\label{eq:app-two-route}
\]
time and $\Os(2^r)$ space. It sorts sources by load, answers complete prefixes
by a shared path dynamic program, and tests residual sources individually.

For each $s\ge2$, take $r=\min\{n,\lfloor3s/2\rfloor\}$. Cover all $s$-sets
by $r$-sets, using $\Os(\Bin ns/\Bin rs)$ members. The construction and its
time bound are proved on page~\pageref{construction}. For each member $U$,
compute the layers in \eqref{eq:app-two-route} on $U$ for $\lceil s/2\rceil\le t<s$, and take the
minimum at each target. If $S\subseteq U$, every partition of $S$ lies
entirely in $U$. Thus a single member containing $S$ already gives its exact
two-route cost. Local tour costs use the original arcs and depot.

When $r=\lfloor3s/2\rfloor$, put $k=\lfloor s/2\rfloor$. Then $r-s=k$ and
$s-t\le k\le r/2$, giving $\Bin r{s-t}\le\Bin rk=\Bin rs$.
Also $r=3k$ or $3k+1$, and $\Bin rk\ge3^k$ by choosing one element from
each of $k$ disjoint triples. Hence $2^{r/2}\le\sqrt2\Bin rs$. The $2^r$
term in \eqref{eq:app-two-route} is absorbed, and the work for this $s$, apart from constructing the
cover, is at most
\[
 \Os\!\left(\frac{\Bin ns}{\Bin rs}\,2^{r/2}\Bin rs\right)
   =\Os\!\left(\Bin ns 2^{r/2}\right)
   \le\Os\!\left(\Bin ns (2^{3/4})^s\right).
 \tag{A.3}\label{eq:app-sum-cover}
\]
Summing \eqref{eq:app-sum-cover} over $s$ gives $\Os((1+2^{3/4})^n)$ by the binomial theorem.

If $n<\lfloor3s/2\rfloor$, use $U=V$. Then $t\ge s/2>n/3$. For the binary
entropy $H$ (base two), concavity and symmetry give
\[
 \sqrt{\Bin n{s-t}\Bin ns}\le
  2^{nH((1-t/n)/2)}\le 2^{nH(1/3)}.
\]
The base of \eqref{eq:app-two-route} in this case is at most
$2^{1/2+H(1/3)}=3/2^{1/6}<2.68$. All polynomially many local layers
return minimizing routes, which can be stored as global masks.

\paragraph{Why the covering cost changes.}
The route-covering construction covers just the added route. For every covering member it
must also enumerate the source customers outside that member. Here the cover
contains the entire target, so all its sources lie inside the member. The
binomial factor for the local targets cancels in \eqref{eq:app-sum-cover}.

\subsection{A deterministic covering construction}\label{construction}
\begin{lemma}
For $0\le s\le r\le n$, a family of $r$-sets containing every $s$-set in at
least one member can be generated deterministically in $\Os(2^n)$ time and
space. Its size, counting repeats, is $\Os(\Bin ns/\Bin rs)$.
\end{lemma}
\begin{proof}
Split the ground set into two fixed halves of sizes $n_1,n_2$ differing by at
most one. Consider a target profile $s_1+s_2=s$, $0\le s_i\le n_i$.
For a uniform random $r$-set, the probability of containing a particular such
target is $p=\Bin rs/\Bin ns$. Conditional on the two half sizes $r_1,r_2$,
the probability is
\[
 p(r_1,r_2)=\prod_{i=1}^2\frac{\Bin{r_i}{s_i}}{\Bin{n_i}{s_i}},
\]
with a zero numerator when $r_i<s_i$. Some feasible allocation has conditional
probability at least $p$ by averaging. Find a maximizing allocation by scanning
the at most $n+1$ choices of $r_1$ and comparing integers.

In each half, greedily cover its $s_i$-sets by $r_i$-sets. With $u$ uncovered
targets, averaging over members shows that one contains at least
$u\Bin{r_i}{s_i}/\Bin{n_i}{s_i}$ of them. Repeating the best choice uses
$O((n+1)\Bin{n_i}{s_i}/\Bin{r_i}{s_i})$ members, including the case $s_i=0$.
Unions of members of the two families cover the target profile. Their number
is $O((n+1)^2/p(r_1,r_2))=O((n+1)^2/p)$. There are at most $n+1$ profiles.

On a half of size $m$, scanning all candidate counters in each greedy iteration
costs $O^*(4^m)$ in total. Initialize each counter to $\Bin{r_i}{s_i}$.
On first covering a target, decrement the counter of every superset of the
prescribed size. All updates together cost $O^*(3^m)$ by counting nested
pairs. Flags and counters cost $O^*(2^m)$ space. Since $m\le\lceil n/2\rceil$,
the construction fits in $O^*(2^n)$ time. The product families may be streamed,
and their total output size is $O^*(1/p)\le O^*(2^n)$.
\end{proof}

\subsection{Small routes in covers with at least three routes}
Use the split evaluation of Lemma~\ref{lem:split}. If $d$ is the suffix size of the
last tour, its two enumeration counts are
\[
 X_d=\frac{n!}{(t-d)!\,(s-t)!\,(n-s+d)!},\qquad
 Y_d=\frac{n!}{d!\,(n-s)!\,(s-d)!}.
\]
The layer costs $\Os(2^n+\min_{0\le d\le t}(X_d+Y_d))$ and returns a witness.
Here $c=s-t\ge2t$. This permits a sharper estimate than one based only on $t/n$.

Put $v=5/3$, $u=v^{-2}=9/25$. The multinomial theorem bounds the counts by
\[
 X_d\le E_1(d)=(1+u+v)^n u^{-(t-d)}v^{-c},\qquad
 Y_d\le E_2(d)=(2+u)^n u^{-d}.
\]
For example, $X_d u^{t-d}v^c$ is one term in $(1+u+v)^n$.
The product is independent of $d$ and
\[
 E_1(d)E_2(d)=((1+u+v)(2+u))^n v^{2t-c}
              \le (13393/1875)^n.
\]
On the real line $E_1$ decreases and $E_2$ increases. At their crossing both
equal the square root of their product. If the crossing lies in $[0,t]$,
rounding down increases $E_1$ by at most $1/u$ and decreases $E_2$.
If it lies below zero, use $d=0$, where $X_0$ is bounded by the square root
and $Y_0=\Bin ns\le2^n$. If it lies above $t$, use $d=t$, where $Y_t$ is
bounded by the square root and $X_t=\Bin n{s-t}\le2^n$. The implementation
can minimize $X_d+Y_d$ exactly over integer $d$. Thus all small-route layers
in \eqref{eq:app-small-route} have base at most $\sqrt{13393/1875}=2.6726267\ldots$.

\paragraph{Completing the algorithm.}
Combine \eqref{eq:app-small-route}--\eqref{eq:app-sum-cover} through budget $\min(K,n)$, taking $K=n$ for an unlimited
fleet. The path tables and all masks recording minimizing routes fit in
$\Os(2^n)$ space, and local computations are sequential. Following the masks
and the Held--Karp predecessors returns actual tours. A zero budget and an
empty customer set are immediate boundary cases. All operations on costs and
loads have bit complexity polynomial in the binary input length. The two
auxiliary bases are below $67/25$, since
$13393/1875<(67/25)^2$ and $729/2<(67/25)^6$.
Finally $(42/25)^4<8$ gives $67/25<1+2^{3/4}$.

\paragraph{How far the count can be sharpened.}
Within the product of the covering ratio and the square-root term in \eqref{eq:app-two-route},
$r=3s/2+O(1)$ is optimal on the equal-route layer $t=s/2$. After dividing
by $\Bin ns$ and squaring, the expression is
$D(r)=2^r\Bin r{s/2}/\Bin rs$. Its successive ratio is
\[
 D(r+1)/D(r)=2(r+1-s)/(r+1-s/2).
\]
It crosses one at $r+1=3s/2$. At the integer minimum, the count agrees with
$\Bin ns 2^{3s/4}$ up to a constant factor before squaring and normalization.
Its largest term occurs at $s/n\to 2^{3/4}/(1+2^{3/4})<2/3$ and attains
base $1+2^{3/4}$, also when $s$ is restricted to even integers. This follows
from Stirling's formula or adjacent binomial ratios. This sharpness statement
concerns the counting estimate. The optimal base for exact CVRP remains open.
For an arbitrary convolution operand the route-removal choice in \eqref{eq:app-small-route} is
unavailable, so this proof covers the indicated layer ranges only.

\paragraph{Reproducibility.}
\texttt{cvrp\_route\_count.py} implements the new schedule using the paper's
path and layer routines. \texttt{verify\_route\_count.py} checks the numerical
comparisons, target covers, two-route values and witnesses, and complete
solutions against direct enumeration.

\section{Three anchors on a directed tour}\label{app:anchors}
Start with a feasible incumbent tour, for example the input vertex order closed at the depot, and denote its cost by $I$. After padding, let $N=3m$ count all vertices and put $h=N-3$. For the general algorithm one can append a forced chain of at most two vertices immediately before the depot, redirect the original incoming depot costs to the first of these vertices, and price disallowed arcs above a known feasible incumbent. Removing the chain recovers the original tour and cost. We consider both cyclic orders on each of the $\binom N3$ anchor sets.

\subsection{Cost filters and the triangle computation}
For each of the three successive ordered pairs of anchors, compute the minimum elementary path cost on each subset with $m-1$ interior vertices. The Held--Karp recursion through size $m-1=h/3$ costs $\Os(2^{H(1/3)h})$, where $H$ is the binary entropy. If a path-set price and the other two table minima sum to at least the incumbent $I$, that set cannot occur in a tour improving $I$. The surviving sets define each ordinary vertex's list of possible segments. An empty list eliminates this anchor order.

For a binary list connect its two segments by an edge bearing that vertex. Singleton lists give fixed memberships. After choosing memberships for the $t$ vertices with three choices, let $a,b,c$ count the edges in the three segment pairs and let $l=\min(a,b,c)$. Choices on these pairs are indexed independently by $i,j,k$; the sum to minimize is $A_{ij}+B_{jk}+C_{ki}$. Partition each dimension into blocks of side $2^l$, do one square min-plus product per block and scan its sum with the third table. With an integer product of exponent $3-\delta$, this costs $\Os(2^{a+b+c-\delta l})$. Alternatively, enumerate the sets of the segment opposite the smallest pair. Each fixes its incident choices, leaving at most $2^l$ completions. This costs $\Os(2^{H(1/3)h+l})$. Take the cheaper procedure. The maximum of its upper exponent over $l$ occurs at the crossing with $h-\delta l$. Over all assignments of the ternary lists the bound is
\[
  \Os\left(3^t 2^{(1-g)h}\right),\qquad
  g=\frac{\delta(1-H(1/3))}{1+\delta}.
\]
Stoian's bit-complexity analysis gives $\delta=1/1944$ deterministically for arbitrary binary integers. The reduction substitutes a sufficiently large finite sentinel for forbidden entries and recovers a witness by scanning the winning block. Thus $t=0$ gives base below $1.999942$ and $t\le h/75000$ gives base below $1.999972$. The reference code accepts an external product for this bound. Its default routine uses a specialized product for few column prices and an ordinary cubic product on general diverse prices. The arbitrary-weight asymptotic speedup is a theorem, and the tests check the exactness of the join. \href{https://arxiv.org/html/2610.10282v2}{Stoian, Appendix A, Lemma 8 and its proof}.

Discard an anchor order if a segment has no surviving set or a vertex has no allowed segment. Evaluate the join above when $t\le h/75000$. Call the remaining orders \emph{unresolved} and retain all three cyclic rotations of each oriented triple. If an improving tour remains after all easy joins, every equally spaced triple on that tour is unresolved, since it was tested at an incumbent at least as large as the final one. For a tour $v_0,\ldots,v_{N-1}$, read columns $(v_j,v_{j+m},v_{j+2m})$ for $j=0,\ldots,m-1$. Guess the first column with its first vertex the depot. For each guessed first column, a state records the visited mask and the last column and stores the least partial cost. Nonnegative arc costs permit pruning a state whose partial cost reaches the current incumbent. Append a disjoint allowed column, paying the three transitions along the rows. Close the last column to the first after cyclically shifting its rows, then read the rows successively. This enumerates exactly the remaining candidate tours, up to taking minima on equal states. Each visited mask is a binary sum of the incidence vectors of the columns. If their rank over $\mathbb F_2$ is $r$, there are at most $2^r$ masks; the other indices contribute polynomial factors. Together the two stages take $\Os(1.999972^N+2^r)$ using the stated product.

\subsection{A metric where every anchor triple survives}
Take a core on $p=18L$ vertices, $L\ge12$, with arbitrary nonnegative directed integer costs $c$. Add $M=3(1+\max c)$ to each off-diagonal core cost to get $w$, setting its diagonal to zero. This makes it a directed metric and shifts every core tour by $pM$. Attach $L$ Petersen graphs at the common depot, adding nine vertices per copy. Their metric $d$ is 1 on an edge and 2 on a distinct nonedge. Let $\pi$ and $\rho_j$ project onto the core and onto copy $j$, respectively, collapsing everything outside a component to its depot. On distinct vertices define
\[
 C_{xy}=1+w_{\pi x,\pi y}+H_0\sum_j d(\rho_jx,\rho_jy),\qquad
 H_0=4(N+1)(1+\max w),\quad N=p+9L .
\]
Set the diagonal to zero. The uniform off-diagonal 1 makes the triangle inequality strict on three distinct vertices. The rooted tour of one entire copy costs 11. The Petersen graph is non-Hamiltonian, while deleting any nonroot vertex leaves a unit-edge cycle on the other nine vertices. These fixed finite facts can also be verified by testing all permutations of the nine nonroot labels. Inserting the removed vertex next to a neighbor in that shorter cycle adds at most two units. Shortcutting any tour on each projection and concatenating component optima at the depot give
\[
 \operatorname{OPT}(C)=N+pM+11LH_0+\operatorname{OPT}(c).
\]

Fix any three ordered anchors, a nonanchor core vertex $v$, and a prescribed segment for $v$. In the relaxation each segment takes a probability distribution on paths with its prescribed endpoints and exactly $m-1$ interior vertices. Over the three distributions each nonanchor vertex must have total marginal 1. We give such distributions containing $v$ on the prescribed segment and of total expected cost strictly below $\operatorname{OPT}(C)-1$.

Let $k\le3$ anchors lie in $u$ distinct copies. In each touched copy, select one primary anchor. Follow a cost-11 rooted tour, shortcutting the other anchors there, and place its two parts immediately before and after the primary anchor in the adjacent segments. Each secondary anchor can be charged a separate depot excursion of cost at most 4. The touched copies thus cost at most $(11u+4(k-u))H_0$ in total. For each segment, leave the untouched copies empty with probability $15/24$. For each possible missing nonroot label, with probability $1/24$ visit every untouched copy on its length-9 cycle omitting that label. Nonroot labels then have marginal $8/24$ in each segment. The expected total cost per untouched copy is $3(9/24)9H_0=81H_0/8$.

Insert $v$ in the chosen segment. Fill each segment to the required size by taking a consecutive interval, from a uniformly random starting point, of a fixed cycle on the remaining unassigned core vertices. The construction occupies at most $8L+1$ interior slots in any segment before filling, whereas $m-1=9L-1$. There are enough free core vertices even to fill a whole segment. The expected total number of filler slots equals the number of free vertices, so each receives total marginal 1. The copy saving below $11LH_0$ is at least $[(L-u)7/8-4(k-u)]H_0\ge13H_0/8$. This exceeds $N\max w+1$. The uniform contribution is $N$. Consequently even the relaxation with the specified position of $v$ beats the incumbent threshold. In particular, some surviving path set for that segment contains $v$: otherwise the expected price with the other segments at their floors would already exceed the threshold. Every nonanchor core vertex therefore has all three positions for every anchor order.

It remains to show that the empty-list test also passes for every nonanchor Petersen vertex $w$. Let $I$ be the incumbent, let $f_r$ be the minimum path-set price on segment $r$, and put $F=\sum_{r=1}^3 f_r$. Draw a path set from each of the three distributions just constructed and write $P_r$ for its price. Then $F\le\sum_r\mathbb E[P_r]<I$. If every set containing $w$ were discarded, the filtering rule would give $P_r-f_r\ge I-F$ whenever the sampled set on segment $r$ contains $w$. All price excesses are nonnegative. Since the three inclusion probabilities for $w$ sum to one,
\[
 \sum_{r=1}^3\mathbb E[P_r-f_r]
 \;\ge\;(I-F)\sum_{r=1}^3\Pr[w\text{ is on segment }r]=I-F.
\]
This would give $\sum_r\mathbb E[P_r]\ge I$, a contradiction. Thus every list is nonempty. At least $18L-3$ vertices have three choices, exceeding the cutoff $h/75000$, so every anchor order is unresolved.

\subsection{The same instance under two algorithms}
Start a column at the depot and two other core vertices. Choose $\lfloor9L/2\rfloor-4$ copy vertices and $j$ of the remaining $p-3$ core vertices, where $j$ is the largest integer at most $\lfloor(p-3)/2\rfloor$ for which the total selected cardinality, including the starting column, is divisible by three. This changes the central index by at most two. Arrange the chosen vertices in three equally long rows, keeping the vertices of each copy together in the concatenated order before cutting it into rows. The row boundaries add at most two copy runs. Visiting $j$ vertices of any copy in one run costs at most $2(j+1)H_0$ including the incident transitions. There are at most $9L/2-4$ selected copy vertices and $L+2$ copy runs, for a contribution of at most $2(9L/2-4+L+2)H_0=(11L-4)H_0$. Core and unit costs add at most $N(1+\max w)<H_0$. Thus the partial cost is strictly below $11LH_0\le\operatorname{OPT}(C)$, and nonnegative arc costs give the same strict inequality for every prefix. The number of distinct selected masks is the product of two near-central binomial coefficients, hence $2^{N-O(\log N)}$. This proves the counting claim for the rotating algorithm. For the savings algorithm, all links between different components have the same depot saving 1. The core savings exceed this by at least $M-\max c$. At threshold zero, the entire core is a connected component separate from every copy, so the independent subproblem has at most $18L-1$ customers. This proves the upper bound in Proposition~\ref{pr:separation}.

The separation survives integer perturbations $S C_{ij}+u_{ij}$ with $S>0$ and $0\le u_{ij}<S/(6N)$. The contribution of the perturbations on any tour, or in expectation on the three segments, is below $S/6$. Cross-component savings vary by at most $3\max u$. At that width after subtracting the global minimum saving, the core component has every internal gain above $(N-2)$ times the width and is therefore certified as a single run. There are at most $9L+1$ resulting blocks and the largest has $18L-1$ customers. When $L$ is kept fixed and the core grows, removing the copies saves only a constant number of vertices. This explains the role of the ratio $p/L$ in the improved exponential base.

\subsection{Three full vehicles with binary choices}
For even $m$, set $B=2m+1$. Three groups of $m$ customers have demands $1,B,B^2$, respectively. Give three anchor customers demands $Q-R_1,Q-R_2,Q-R_3$, where
\[
 (R_1,R_2,R_3)=\tfrac m2(1+B^2,\,1+B,\,B+B^2),\qquad Q=2\max_iR_i+1 .
\]
With at most three vehicles the anchors must be separated and every vehicle is full. Uniqueness of the digits without carries forces exactly half of each eligible group on each of its two vehicles. There are $\binom m{m/2}^3=2^{n-O(\log n)}$ feasible assignments, $n=3m+3$. Each local price table takes $\Os(2^{2m})$ to build. A triangle join of dimension $2^m$ then gives a deterministic $\Os((2^{1-1/5832})^n)$ upper bound analytically, with the same integer product as above. The construction imposes no numerical restriction on directed arc prices.